\PassOptionsToPackage{table}{xcolor}
\documentclass[sigconf,nonacm]{acmart}
\AtBeginDocument{%
  }

\setcopyright{none}
\copyrightyear{2026}
\acmYear{2027}
\acmDOI{}
\acmConference{}{}{}
\acmISBN{}

\usepackage{array}
\usepackage{makecell}
\usepackage{pifont}
\usepackage{listings}
\usepackage{tabularx}

\lstdefinestyle{yamlconfig}{
    basicstyle=\ttfamily,
    columns=fullflexible,
    keepspaces=true,
    breaklines=true,
    frame=single,
    rulecolor=\color{black!25},
    backgroundcolor=\color{black!3},
    xleftmargin=0.5em,
    xrightmargin=0.5em,
    framexleftmargin=0.5em,
    framexrightmargin=0.5em
}

\newcommand{\yes}{\ding{108}}   % filled circle
\newcommand{\maybe}{\ding{119}} % half-ish/outlined marker fallback
\newcommand{\no}{\ding{109}}    % open circle

\usepackage{xspace}
\newcommand{\systemname}{\texttt{CREST}\xspace}

\usepackage{xcolor}

\usepackage{tikz}

\usepackage[table]{xcolor}

\usepackage{stfloats}
\usepackage{placeins}
\graphicspath{{assets/}}

\begin{document}

\title{Measure the Deployment, Not Just the Model: Rethinking Hardware-Aware Model Selection for Microcontrollers}

% \author{Anonymous Authors}
\author{Joseph Q. Zales}
\affiliation{%
  \institution{University of California, Los Angeles}
  \city{Los Angeles}
  \state{California}
  \country{USA}
}
\email{jzales@ucla.edu}

\author{Pragya Sharma}
\affiliation{%
  \institution{University of California, Los Angeles}
  \city{Los Angeles}
  \state{California}
  \country{USA}
}
\email{pragyasharma@ucla.edu}

\author{Mani Srivastava}
\affiliation{%
  \institution{University of California, Los Angeles}
  \city{Los Angeles}
  \state{California}
  \country{USA}
}
\email{mbs@ucla.edu}

% \renewcommand{\shortauthors}{Zales and Srivastava}

\begin{abstract}
Selecting an energy-efficient model for periodic sensing requires accounting for the time between inferences. Hardware-aware model search typically measures inference alone, even when the application cares about energy per release. We present \systemname{} (Cross-platform Runtime Evaluation and Search Tool), an open-source modular hardware-in-the-loop (HIL) experimentation system that evaluates model and hardware configurations under a specified release schedule. \systemname{} lets researchers vary the selection objective, runtime schedule, target and candidate proposer while reusing the evaluation and replay machinery. Its measurement records preserve the conditions needed to compare results and distinguish valid measurements from deployable candidates. Three case studies demonstrate this capability and the deployment insights it enables. First, six searches on STM32 N657 compare continuous-energy and periodic-energy selection at release periods of 100, 200 and 500~ms. Controlled replay of configurations from these searches shows how waiting changes selection. For one architecture, continuous inference favors 800~MHz over 300~MHz at 8.0 versus 17.1~mJ per inference. A fitted decomposition shows that additional waiting offsets 97\% of this saving. With STOP between releases, 300~MHz uses about 0.3--0.4\% less board energy per release at 100 and 500~ms. At a fixed 300~MHz clock, the preferred architecture changes between the 100 and 500~ms release periods, which a model with shared waiting parameters cannot produce. Across held-out search observations, latency predicts continuous energy within a median 1.7\%, yet most release-energy error remains in the scheduled terms. In paired replay, one of six randomly selected near-latency pairs reverses its continuous-energy ordering by 5.8~mJ per release. Second, we compare proxy-selected and measured-energy selections for inertial odometry on three Arduino target configurations. Target replay shows that proxy-selected candidates can fail to fit or use more energy, with only 9 of 16 FLOPs-selected candidates fitting on the most memory-constrained board. Third, an audio classification study reuses the same evaluator with a new workload and LLM-based candidate proposers, and cross-target replay shows a selected configuration moving from third to last in energy rank on another board.
\end{abstract}

%%
%% FIXME: DO this again after the paper is written
\begin{CCSXML}
<ccs2012>
   <concept>
       <concept_id>10010520.10010553.10010562.10010564</concept_id>
       <concept_desc>Computer systems organization~Embedded software</concept_desc>
       <concept_significance>300</concept_significance>
       </concept>
   <concept>
       <concept_id>10010520.10010570.10010573</concept_id>
       <concept_desc>Computer systems organization~Real-time system specification</concept_desc>
       <concept_significance>100</concept_significance>
       </concept>
   <concept>
       <concept_id>10010583.10010682.10010696</concept_id>
       <concept_desc>Hardware~Modeling and parameter extraction</concept_desc>
       <concept_significance>300</concept_significance>
       </concept>
   <concept>
       <concept_id>10011007.10011006.10011072</concept_id>
       <concept_desc>Software and its engineering~Software libraries and repositories</concept_desc>
       <concept_significance>300</concept_significance>
       </concept>
 </ccs2012>
\end{CCSXML}

\ccsdesc[500]{Computer systems organization~Embedded software}
\ccsdesc[500]{Computing methodologies~Neural networks}
\ccsdesc[300]{Computing methodologies~Machine learning}
\ccsdesc[300]{Software and its engineering~Software libraries}

%%
%% FIXME: Create keywords here
\keywords{Neural architecture search, TinyML, hardware-in-the-loop, embedded machine learning, energy measurement}

% FIXME: Update this section too

\maketitle

\section{Introduction}
\label{sec:intro}

Resource-constrained microcontrollers (MCUs) are increasingly the deployment target for neural network inference in sensing applications, spanning workloads such as inertial odometry~\cite{saha_tinyodom_2022}, vibrometry~\cite{kimura_vibrofm_2024}, digital biomarkers~\cite{hossain_mcerebrum_2017}, audio classification~\cite{nordby_environmental_2019}, and keyword spotting~\cite{zhang_hello_2018,bartoli_end--end_2025}. Embedded inference runtimes such as TensorFlow Lite Micro~\cite{david_tensorflow_2021} and optimized kernels such as CMSIS-NN~\cite{lai_cmsis-nn_2018}, alongside recent Cortex-M cores with low-precision arithmetic support~\cite{arm_helium,arm_cortex_m55}, have made neural-network inference feasible within hundreds-of-KB to MB-scale MCU SRAM~\cite{arduino_nano33ble_sense_datasheet,arduino_portenta_h7_datasheet,st_stm32n6_datasheet}. However, selecting a model requires accounting for the memory, latency, and energy constraints of its deployment. The same model architecture can satisfy these constraints under one deployment configuration and violate them under another.

Hardware-in-the-loop (HIL) neural architecture search (NAS) systems use target measurements to guide model selection~\cite{saha_tinyodom_2022,saha_tinyns_2024,groh_edgevolution_2026}. Benchmarks and deployment tools support comparisons across workloads and targets~\cite{banbury_mlperf_2021,hasanpour_edgemark_2025,van_kempen_mlonmcu_2023}. Inference measurements and calibrated estimates provide useful costs for model selection. Their value depends on whether they preserve the choice under the intended schedule and target. A reusable hardware evaluation path lets researchers check these choices and reevaluate candidates when deployment conditions change, without rebuilding the measurement and selection workflow.

This paper presents \systemname{} (Cross-platform Runtime Evaluation and Search Tool), a modular HIL experimentation system for embedded model selection. Researchers can vary the objective, schedule, target and candidate proposer while reusing evaluation, logging and replay. The same evaluator supports measurements during search and replay of candidates selected by other methods. Continuous execution repeats inference without an intentional waiting interval and reports energy per inference. Periodic execution releases one inference every specified period and reports energy per release, including inference and the following sleep or waiting interval.

Each evaluation produces a measurement record that identifies its interval and target and separates measurement validity from deployment feasibility. A candidate that misses its deadline can therefore retain a valid energy measurement while being excluded from selection. Replay measures selected configurations on another target while keeping source results separate from destination measurements. Workload and proposer interfaces let researchers change how candidates are built and chosen while retaining the evaluator. Table~\ref{tab:related-work-capabilities} compares these capabilities with related systems.

Using this system, we investigate when inference-based model and clock choices transfer to periodic execution. Both waiting energy and the release period can change the preferred configuration. We test whether calibrated estimates preserve these choices and use scheduled replay to identify differences they miss. Calibrated predictors can guide selection; \systemname{} provides a reusable hardware evaluation path to check whether the resulting choice meets deployment constraints and retains its energy advantage under the intended target and schedule.

On STM32 N657, a fitted decomposition shows that waiting offsets 97\% of a faster clock's inference saving. At a fixed clock, the preferred architecture changes with the release period, which shared waiting parameters cannot represent. Latency predicts continuous energy well, but scheduled replay exposes differences missed by the evaluated estimates. Target replay exposes proxy-selected deployment failures and cross-board energy-rank changes.

Three case studies evaluate these contributions across two workloads and three Arm Cortex-M boards. CS1 investigates schedule-dependent model and clock selection on STM32 N657 and evaluates calibrated estimates on held-out architectures and repeated measurements. CS2 and CS3 examine target-dependent deployment choices while reusing the evaluator across workloads and proposers. CS2 compares proxy-selected odometry candidates with measured-energy selections through target deployment and replay. CS3 adds audio classification and LLM-based proposers to the shared evaluator, then replays selected configurations across targets. Measurements are taken at the board supply.

This work makes three contributions.
\begin{itemize}
    \item \textbf{Evidence that deployment schedules can change selection.} Controlled experiments show that waiting can offset most of an inference saving and change the preferred clock. At a fixed clock, the preferred architecture can change with the release period, which shared waiting parameters cannot represent.
    \item \textbf{An evaluation of calibrated energy estimates.} Held-out tests show that latency predicts continuous energy well and that estimating this term usually preserves selection. Most release-energy error lies in the scheduled terms, and paired replay identifies scheduled differences that the estimates miss.
    \item \textbf{Reusable evaluation and replay infrastructure.} These findings are supported by \systemname{}, an open-source HIL system with modular evaluation, measurement records and replay across workloads, targets and candidate proposers\footnote{The \systemname{} repository is available at \url{https://github.com/nesl/crest}.}.
\end{itemize}

\begin{table*}[htbp]
\centering
\caption{Capabilities of \systemname{} and related systems. \yes{} denotes central support, \maybe{} partial support and \no{} absent or noncentral support. Search covers model, program, or configuration choices, not buffer or compiler tuning. Proxy and HIL costs must guide model selection. HIL requires live candidate measurements, not offline profiling or post-search validation. Measured targets requires a common method or workload measured on at least two physical targets. Extensibility requires reusable workload/model or backend interfaces. Model type describes the central optimized model/program. N/A denotes non-model-centered systems.}
\label{tab:related-work-capabilities}
\scriptsize
\setlength{\tabcolsep}{3pt}
\renewcommand{\arraystretch}{1.12}
\begin{tabular}{>{\raggedright\arraybackslash}p{2.65cm} p{1.2cm} p{1.35cm} p{2.15cm} p{2.0cm} c c c c c c}
\toprule
Work &
\makecell{Artifact\\type} &
\makecell{Model\\type} &

\makecell{Target\\scope} &
Availability &
\makecell{Automated\\model/design\\search} &
\makecell{Measured\\targets} &
\makecell{Proxy-based\\model selection} &
\makecell{HIL model\\selection} &
\makecell{Extensible\\tasks/targets} &
\makecell{Schedule\\options} \\
\midrule
\rowcolor{cyan!10}  % pale yellow
\textbf{\systemname{}}& \textbf{Full tool} & \textbf{Neural} & \textbf{Multi-vendor MCUs} & \textbf{Open source} & \textbf{\yes} & \textbf{\yes} & \textbf{\yes} & \textbf{\yes} & \textbf{\yes} & \textbf{\yes} \\
TinyNS 2024 \cite{saha_tinyns_2024} & Full tool & Neurosymbolic & STM32 family & Open source & \yes & \yes & \yes & \yes & \yes & \no \\
TinyOdom 2022 \cite{saha_tinyodom_2022} & Full tool & Neural & STM32 family & Open source & \yes & \yes & \yes & \yes & \no & \no \\
EdgeVolution 2026 \cite{groh_edgevolution_2026} & Full tool & Neural & Zephyr RTOS & Open source & \yes & \yes & \maybe & \yes & \yes & \no \\
Edge Impulse 2023 \cite{hymel_edge_2023} & Product & Neural & MCU + CPU & Product & \yes & \yes & \yes & \no & \yes & \no \\
MicroNAS 2025 \cite{king_micronas_2025} & Search stage & Neural & STM32 family & Available on request & \yes & \yes & \yes & \no & \no & \no \\
MicroNets 2021 \cite{banbury_micronets_2021} & Search stage & Neural & STM32 family & Open source & \yes & \yes & \yes & \no & \no & \no \\
$\mu$NAS 2020 \cite{liberis_nas_2020} & Search stage & Neural & Single MCU & Open source & \yes & \no & \yes & \no & \no & \no \\
TinyTNAS 2024 \cite{saha_tinytnas_2024} & Search stage & Neural & Multi-vendor MCUs & Open source & \yes & \yes & \yes & \no & \no & \no \\
ReactQuant 2025 \cite{misra_latency-constrained_2025} & Search stage & Neural & Single MCU & No public artifact found & \yes & \no & \yes & \no & \no & \no \\
\mbox{Embedded Arena 2026~\cite{zhang_embedded_2026}} & Optimizer & N/A & Multi-vendor MCUs & Open source & \yes & \yes & \no & \yes & \yes & \no \\
EdgeMark 2025 \cite{hasanpour_edgemark_2025} & Benchmark & N/A & Multi-vendor MCUs & Open source & \no & \yes & \no & \no & \yes & \no \\
MLonMCU 2023 \cite{van_kempen_mlonmcu_2023} & Benchmark & N/A & Multi-vendor MCUs & Open source & \no & \yes & \no & \no & \yes & \no \\
HW-NAS-Bench 2021 \cite{li_hw-nas-benchhardware-aware_2021} & Benchmark & N/A & CPU + accelerator & Open source & \yes & \yes & \yes & \no & \no & \no \\
InstMeter 2026 \cite{liu_instmeter_2026} & Estimator & N/A & Multi-vendor MCUs & No public artifact found & \no & \yes & \yes & \no & \no & \no \\
\bottomrule
\end{tabular}
\end{table*}

\section{Related Work}
\label{sec:related-work}

\systemname{} draws on hardware-aware search and reusable deployment tools, whose capabilities are compared in Table~\ref{tab:related-work-capabilities}. Work on periodic sensing and frequency scaling provides the foundation for our study of how deployment schedules affect model and clock selection.

\textbf{Proxy-based MCU NAS.} MCU NAS often uses inexpensive objectives such as MACs, operation counts, RAM, flash, or latency lookup tables~\cite{liberis_nas_2020,lin_mcunet_2020,king_micronas_2025,saha_tinytnas_2024,qiao_monas_2025}. Quantization frameworks such as ReactQuant also select configurations under latency and memory bounds on a single MCU~\cite{misra_latency-constrained_2025}. MicroNets supports this approach with measurements showing that operation count is a useful proxy for latency and energy within its evaluated search spaces and hardware/software settings~\cite{banbury_micronets_2021}. These proxies are useful for pruning search, but prior MCU studies show that kernels, operator mix, memory behavior, and framework choices can change latency and energy enough to make proxy-only search decisions unreliable~\cite{lai_not_2018,heim_measuring_2021,lai_cmsis-nn_2018}.

\textbf{Predictors and HIL measurement.} Energy and latency predictors use calibration data, program behavior or runtime counters to estimate costs without measuring every candidate~\cite{sinha_energy_2000,yuwen_sun_low-cost_2012}. InstMeter is a recent MCU predictor that reports substantially lower energy and latency error than MAC-based and nn-Meter~\cite{zhang_nn-meter_2021} baselines, with approximately 30\% 90th-percentile relative error in its main energy evaluation using five training samples~\cite{liu_instmeter_2026}. Its inference-cost predictions address a different quantity from energy per scheduled release, which includes the waiting interval. Hardware evaluation also checks whether a candidate deploys and meets its timing constraints. Prediction and hardware validation can therefore serve complementary roles in embedded model search.

\textbf{HIL measurement and benchmarking infrastructure.} HIL NAS uses target measurements during model selection. TinyOdom applies hardware-aware search and TCN models to neural inertial odometry, measuring memory and latency on the target~\cite{saha_tinyodom_2022}. CS1 and CS2 study the same application using OxIOD and a TCN model family. \systemname{} additionally evaluates measured energy and periodic execution, allowing the selection objective to include the interval between inferences. TinyNS jointly searches neural and symbolic components and queries target devices to check deployability, memory use and latency across multiple workloads and targets~\cite{saha_tinyns_2024}. EdgeVolution provides open-source HIL NAS across four classification workloads and three Nordic MCUs, with measured inference latency and energy and interchangeable search strategies~\cite{groh_edgevolution_2026}. These systems establish direct target interaction, resource-aware selection and reusable evaluation as prior work. Where EdgeVolution evaluates inference-level hardware costs, \systemname{} also measures energy per release. Its measurement records separate validity from deployment feasibility and preserve intervals and source/target provenance through scoring and replay (\S\ref{sec:system}).

Benchmarking systems make embedded ML results comparable across tasks, tools, and hardware, but they serve a different role than NAS systems that use measurements to select architectures. MLPerf Tiny~\cite{banbury_mlperf_2021} standardizes benchmark tasks, metrics, and reporting requirements for embedded inference. MLonMCU~\cite{van_kempen_mlonmcu_2023} and EdgeMark~\cite{hasanpour_edgemark_2025} extend this toward multi-board characterization with retargetable build flows. HW-NAS-Bench~\cite{li_hw-nas-benchhardware-aware_2021} provides measured and estimated hardware costs as a substrate for NAS algorithm comparison. Edge Impulse~\cite{hymel_edge_2023} is an MLOps platform that supports model tuning, multi-board deployment and dataset management. \systemname{} connects live candidate deployment to task-quality evaluation and selection under configured runtime conditions.

\textbf{LLM-guided embedded optimization.} OPRO demonstrates iterative LLM optimization using previous solutions and their objective values as feedback~\cite{yang_large_2024}. LLMs also support architecture generation and embedded optimization. MONAQ generates architectures for time-series tasks and evaluates MCU resource requirements through simulation~\cite{trirat_monaq_2025}. AutoMCU incorporates vendor-toolchain feasibility checks into LLM-guided network customization~\cite{dai_automcu_2026}. Embedded Arena uses compilation, flashing, and physical-device feedback to guide model and firmware optimization~\cite{zhang_embedded_2026}. \systemname{} separates candidate proposal from evaluation, so LLM proposals and native sampling use the same deployment checks, scoring and measurement records. CS3 demonstrates this reuse with TPE and LLM proposals on two MCU targets.

\textbf{Schedule-aware sensing and frequency scaling.} Many embedded sensing applications operate periodically, making duty cycle and measurement scope important to evaluation. Keyword spotting systems account for frame periods when reporting energy and latency~\cite{zhang_hello_2018,bartoli_end--end_2025}, and acoustic-scene classifiers report measurements relative to duty cycle on low-power MCU devices~\cite{naccari_embedded_2020}. MicroNets~\cite{banbury_micronets_2021} introduces a workload-uptime metric capturing the fraction of input stride spent in active inference. Bartoli et al.~\cite{bartoli_benchmarking_2025} use trigger signals to separate inference from surrounding work in phase-aware energy and latency measurements.

Frequency-scaling studies explain why the energy benefit of finishing early depends on the power drawn while waiting. Miyoshi et al. compare a fast run padded with idle time against a slower run over the same period, showing that the better choice depends on idle power relative to active power~\cite{miyoshi_critical_2002}. Le Sueur and Heiser examine DVFS with idle energy included and show that better idle modes favor running fast~\cite{le_sueur_dynamic_2010}. Kim et al. formulate performance-constrained resource allocation and compare race-to-idle, pace-to-idle and no-idle strategies across four multicore systems, showing that race-to-idle can be substantially suboptimal~\cite{kim_racing_2015}. These studies vary hardware resource settings for fixed benchmark programs, whereas our study compares neural architectures as well as clock settings. Building on these foundations, CS1 tests when inference-based model and clock choices transfer to periodic execution and where calibrated estimates of energy per release miss scheduled costs. \systemname{} provides the scheduled measurement and replay path for this investigation.

\section{Energy per Release}
\label{sec:energy}

\systemname{} evaluates model and deployment choices under a specified schedule. We first describe the energy objective and how \systemname{} measures it. \S\ref{sec:system} then describes the interfaces and records used to evaluate candidates.

\subsection{Continuous and Periodic Energy}
\label{subsec:runtime-schedules}

\systemname{} treats runtime schedule as part of the study definition because sensing deployments usually run as scheduled windows rather than saturated inference loops~\cite{zhang_hello_2018,banbury_micronets_2021,bartoli_benchmarking_2025,bartoli_end--end_2025}. An application requests a prediction every $T$ seconds. A candidate must finish inference within that period, and the board can sleep or wait for the remaining time.

\begin{figure}[t]
    \centering
    \includegraphics[width=\columnwidth]{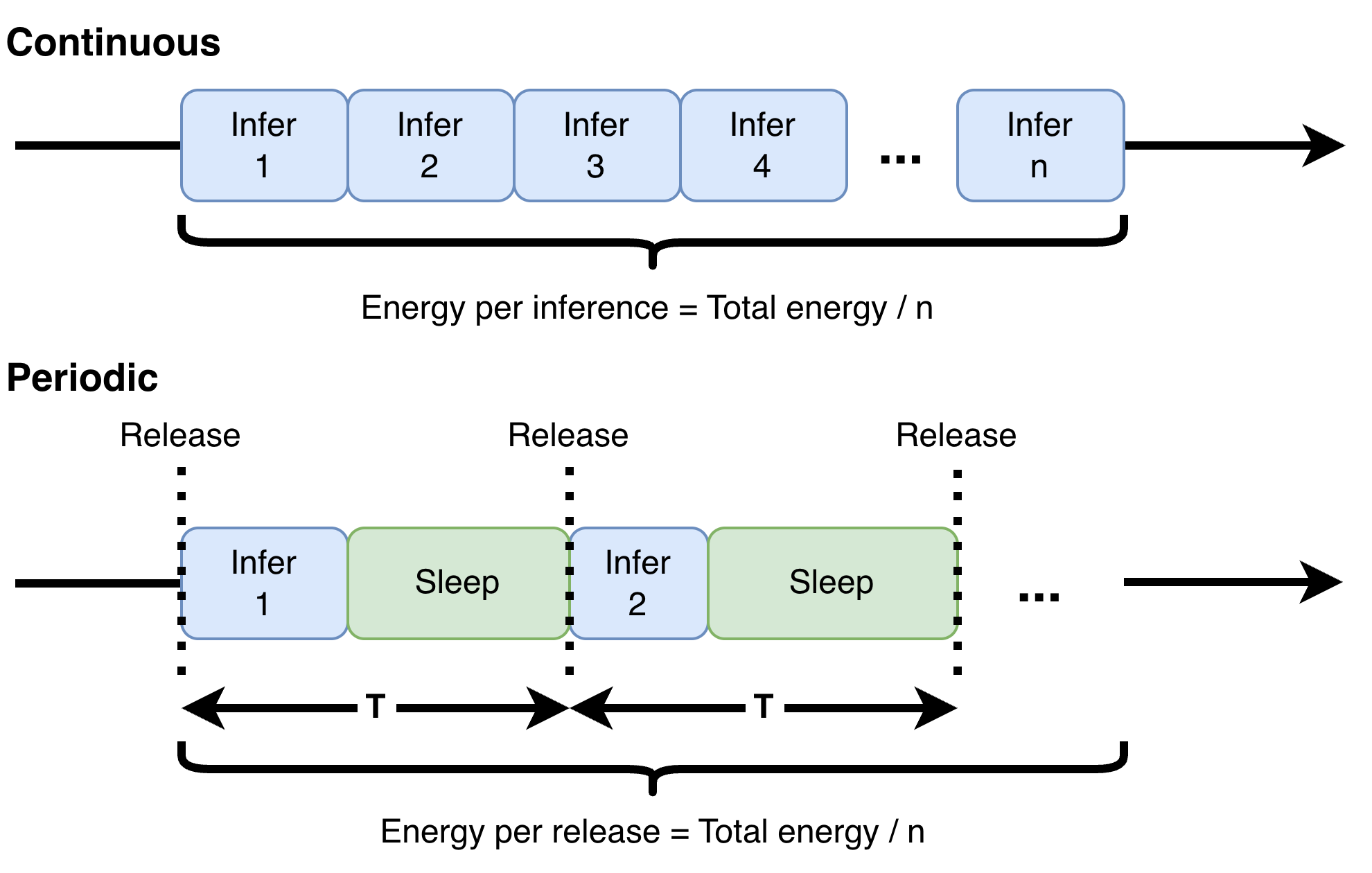}
    \caption{Runtime schedules in \systemname{}. Continuous inference repeats inference without an intentional release interval. Periodic inference releases one inference per period \(T\) and measures each period, including active inference and the following sleep or waiting interval.}
    \Description{A timing diagram with continuous inference as adjacent inference blocks and periodic inference as one inference block per period followed by a sleep or waiting interval.}
    \vspace{-1.5em}
    \label{fig:crest-runtime-modes}
\end{figure}

\systemname{} supports two runtime modes, and the choice changes what ``efficient'' means. \emph{Continuous inference} repeats inference with no intentional release interval and reports energy per inference under a saturated loop. Prior HIL search systems measure inference in this way~\cite{saha_tinyodom_2022,groh_edgevolution_2026}. \emph{Periodic inference} releases one inference every period \(T\) and reports energy per release. After inference completes, the board sleeps when slack permits. With little time before the next release, it may wait awake instead. Figure~\ref{fig:crest-runtime-modes} shows both modes. CS2 and CS3 select models by continuous energy, and CS1 compares the two modes.

To examine when a continuous-energy choice remains preferable under periodic execution, we relate the energy measured in these two modes. Let candidate $x=(m,q,f)$ specify the model configuration $m$, quantization $q$ and CPU frequency $f$, with $f_x$ denoting the frequency component of candidate $x$. For a fixed board and backend, let $C(x)$ be its continuous inference duration and $P_{\mathrm{cont}}(x)$ its mean power over that interval. Its continuous energy per inference is
\begin{equation}
E_{\mathrm{cont}}(x)=C(x)P_{\mathrm{cont}}(x).
\label{eq:crest-continuous-energy}
\end{equation}
The measured objective is board energy over the scheduled window, normalized per release as in Equation~\ref{eq:crest-energy}. To estimate it from continuous measurements, we use the approximation
\begin{equation}
\begin{split}
E(x,T)\approx{}&P_{\mathrm{cont}}(x)C(x)\\
 &+(T-C(x))P_{\mathrm{wait}}(x)+b(x).
\end{split}
\label{eq:crest-release-accounting}
\end{equation}
The first term is power times inference duration. The second is waiting power times the remaining duration. Here $P_{\mathrm{wait}}(x)$ is the effective waiting power under the chosen policy. The correction $b(x)$ accounts for transitions, window effects and differences between continuous inference and inference after wake. We use continuous duration as the timing reference for this approximation. Deployment feasibility is checked using the timing and deadline telemetry of the scheduled execution. With durations in seconds and powers in mW, both products and the per-release correction $b(x)$ are in mJ.

When continuous cost and correction terms are independent of the period and waiting power is shared, the energy difference between two feasible candidates does not depend on the period. Configuration-dependent waiting behavior or period-dependent execution costs can instead change the preferred configuration.

Equation~\ref{eq:crest-release-accounting} permits waiting behavior that depends on the full configuration $x$. Our continuous-derived predictors restrict waiting power and correction to clock-specific values $\widehat P_{\mathrm{wait}}(f)$ and $\widehat b(f)$, shared by all architectures at clock $f$. The measured-energy predictor obtains $P_{\mathrm{cont}}(x)$ by dividing measured continuous energy by $C(x)$
\begin{equation}
\begin{split}
\widehat E_{\mathrm{meas}}(x,T)={}&P_{\mathrm{cont}}(x)C(x)\\
 &+(T-C(x))\widehat P_{\mathrm{wait}}(f_x)+\widehat b(f_x).
\end{split}
\label{eq:crest-window-energy-current}
\end{equation}
Appendix~\ref{app:cs1-waiting-model} gives the waiting calibration.

For the latency-based estimate, we fit continuous energy as an affine function of duration
\begin{equation}
\widehat E_{\mathrm{cont,aff}}(x)=a_{f,q}+s_{f,q}C(x).
\label{eq:crest-affine-continuous}
\end{equation}
The fitted slope $s_{f,q}$ has power units and the intercept $a_{f,q}$ has energy units. Substituting this energy estimate for $P_{\mathrm{cont}}(x)C(x)$ in Equation~\ref{eq:crest-window-energy-current} retains the $(T-C)$ waiting term. A second affine estimate fits energy per release directly from continuous duration under a fixed period and policy. These estimates use calibration architectures excluded from evaluation. The direct estimate requires scheduled-energy calibration measurements. CS1 compares their prediction and selection behavior.

At a fixed frequency, quantization and period, the affine predictor orders candidates by latency when $s_{f,q}>\widehat P_{\mathrm{wait}}(f)$. The measured-energy predictor can give a different ordering when mean continuous power varies across candidates. Because both predictors share waiting parameters at each frequency, they do not represent a period-dependent crossing between candidates at the same frequency and quantization. A direct scheduled fit is calibrated for one period and does not predict crossings across periods.

When both candidates meet the deadline, the faster candidate leaves more time for waiting before the next release. Its continuous-energy saving must exceed the cost of that additional waiting to retain its advantage per release under shared waiting power and corrections. This tradeoff parallels the critical power slope studied in frequency scaling~\cite{miyoshi_critical_2002}. More generally, performance-constrained resource allocation must account for both active and idle energy, and race-to-idle need not minimize their sum~\cite{kim_racing_2015}. The direction of this tradeoff depends on waiting power relative to active power at the measured boundary. Lower sleep power favors finishing early~\cite{le_sueur_dynamic_2010}, so the preferred clock must be evaluated at the deployment's own power boundary. A background load that remains constant throughout the release period contributes the same energy to every candidate, whereas continuous measurements charge slower candidates for that load over a longer inference interval.

\subsection{Measurement Boundary}
\label{subsec:measurement-boundary}

A sensing window may include sampling, feature processing, communication, inference and waiting~\cite{bartoli_benchmarking_2025,schmid_interaction_2010}. Our measurements cover inference and the following waiting interval at the board supply, including all components powered through that boundary (\S\ref{subsec:impl-hil-harness}). We integrate power over the complete measured interval rather than treating waiting time as free.
\begin{equation}
  E_{\mathrm{norm}} =
  \frac{1000}{N}
  \int_{t_0}^{t_1} P_{\mathrm{board}}(t)\,dt
  \label{eq:crest-energy}
\end{equation}
Here, \(P_{\mathrm{board}}(t)\) is board power in watts, \(N\) counts completed invocations in continuous mode or releases in periodic mode, and 1000 converts joules to millijoules.

\section{System Architecture}
\label{sec:system}

\begin{figure*}[t]
    \centering
    \includegraphics[width=0.85\textwidth]{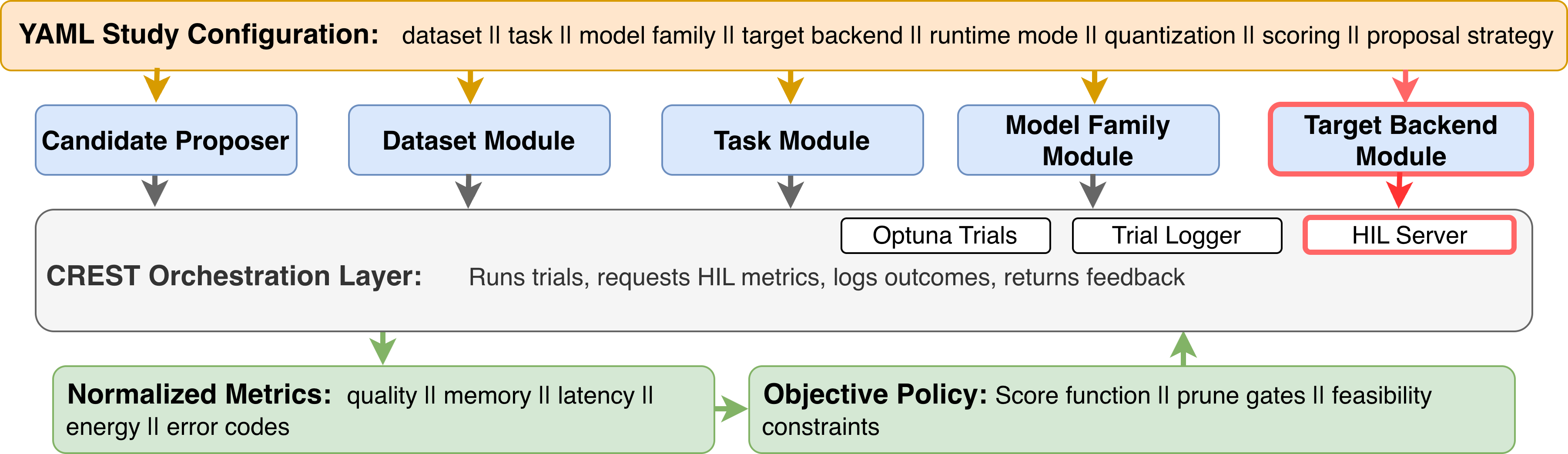}
    \caption{\systemname{} framework structure. Study configuration selects a proposal strategy alongside the dataset, task, model family, target backend, runtime mode, quantization, and scoring policy. TPE, NSGA-II, or LLM proposals enter Optuna-managed trials. Shared orchestration evaluates candidates, requests HIL measurements when needed, and logs normalized metrics. The objective policy returns objective and feasibility feedback.}
    \Description{A stack diagram with a YAML study configuration above five modules named Candidate Proposer, Dataset, Task, Model Family, and Target Backend. They feed into a \systemname{} orchestration layer containing Optuna Trials, Trial Logger, and HIL Server. The layer produces normalized quality, memory, latency, energy, and error-code metrics. An Objective Policy uses those metrics for scoring, pruning, and feasibility checks, then returns feedback to the orchestration layer.}
    \vspace{-1.5em}
    \label{fig:crest-overview}
\end{figure*}

\systemname{} separates candidate proposal, workload evaluation and target measurement through replaceable interfaces. Each trial evaluates a proposed model and deployment configuration, then returns metrics and feasibility information to the search.

Throughout the paper, a \emph{study} refers to one configured NAS optimization. A study begins with choosing the deployment conditions the user wants to test. The \emph{workload} determines the data and prediction task, the \emph{model family} defines the architectures available to the search, the \emph{target backend} sets the board and deployment path, and the \emph{runtime schedule} selects continuous or periodic inference (\S\ref{subsec:runtime-schedules}). The study also includes a \emph{policy}, which names the metrics and rules used for scoring, pruning, and feasibility.

During the study, each optimizer step is a \emph{trial}. A trial evaluates one \emph{candidate}, which is a concrete model and deployment choice. In the current system, a candidate may include model hyperparameters, quantization mode, and backend-supported options such as CPU clock. A trial produces normalized \emph{metrics}, including task quality, proxy cost, memory, latency, energy, status, and schedule telemetry when available. The policy converts these metrics into optimizer feedback. \emph{Feasibility} records whether the candidate satisfies constraints such as memory capacity, latency, deadline, or user-defined limits. A candidate can therefore be measured successfully and still be infeasible.

\systemname{} supports these studies through replaceable modules with explicit interfaces, shown in Figure~\ref{fig:crest-overview}. Optuna-native samplers, LLMs or other proposers implementing \texttt{OptimizerABC} supply candidates, while Optuna manages trial execution and storage~\cite{akiba_optuna_2019}. The workload-facing interfaces own data preparation, task semantics, and model construction. The target-backend interface owns vendor-specific compilation, programming, telemetry, and power measurement. Runtime schedule is also part of the study configuration, so continuous and periodic execution can be selected without changing the workload or model family. A new workload, board, or schedule only changes the relevant module or configuration field, allowing the surrounding evaluation and logging code to be reused. Studies that do not request hardware measurements can use host metrics and compilation reports through the same path. The current modules support two workload/model-family pairs, three MCU backends, continuous and periodic execution, float32 and int8 PTQ, and proxy or measured objectives, with native Optuna sampling and the Memory and History LLM policies providing three proposal schemes.

\subsection{Changing Workloads and Targets}
\label{subsec:adding-workloads-models}

The workload interfaces in \systemname{} separate the dataset, task and model family. These pieces can be replaced together or independently. The dataset defines examples and metadata. The task defines what the model should predict and how quality is measured. The model family defines the model architecture search space and builds each candidate. After that, candidates from the new modules follow the same \systemname{} workflow as built-in candidates.

Users do not always need to replace the whole workload stack. If the task and model family stay fixed, adding a dataset only requires a dataset adapter that provides examples, splits, input shape, data type, calibration data, and optional schedule metadata. If the dataset and task stay fixed, adding a model family only requires a model-family implementation that defines the search space, builds candidates, and exports a model compatible with the task output. A full new workload usually adds a dataset adapter, a task definition and a model family.

If the dataset, task and model family stay fixed, changing to a supported target only requires selecting its backend in the study configuration. Adding a new target requires implementing its toolchain, programming and measurement operations through the backend interface. The backend reports the metrics it supports and marks unavailable fields explicitly. This lets the surrounding search and analysis code use a common record without assuming that every board exposes the same telemetry.

\subsection{Measurement Records}
\label{subsec:measurement-records}

The interfaces above let a study change its proposer, workload, target or schedule. Comparing measurements requires knowing the conditions under which they were collected. \systemname{} returns every trial as a \emph{measurement record}. The record carries the normalized metrics and describes four properties of how they were obtained.
\begin{itemize}
    \item \textbf{Interval.} The runtime mode, release period and invocation count that define the measured window. Continuous energy is reported per completed invocation and periodic energy per release (Equation~\ref{eq:crest-energy}).
    \item \textbf{Validity.} Whether the backend accepted the measurement. Unsupported fields are marked unavailable, and failed checks are retained with the attempt that produced them. On coordinated DUT-plus-harness paths, energy is accepted only when DUT timing exists and DUT and harness run counts match (\S\ref{subsec:impl-hil-harness}).
    \item \textbf{Feasibility.} Whether a valid measurement satisfies the configured memory, latency, deadline and user-defined limits. Validity and feasibility are recorded separately, so a measured deadline miss remains an infeasible result rather than a failed measurement.
    \item \textbf{Provenance.} The study configuration, candidate payload and target that produced each value. Replay keeps inherited source metrics apart from destination measurements (\S\ref{subsec:impl-logging-replay}).
\end{itemize}
The backend checks measurement validity using the telemetry available on its target. The policy applies the study's feasibility rules and selects the metrics used for scoring. Logging retains the measurements and their status, while Pareto export identifies candidates that meet the configured constraints. These consumers use the same record, so selection and later analysis can refer to the same measurement conditions.

\systemname{} also supports replaying candidate configurations to request new measurements through the same backend interface. It keeps destination hardware costs separate from inherited source quality, allowing comparisons across targets while preserving where each value was obtained.

\section{Implementation}
\label{sec:technical-implementation}

\begin{figure}[t]
    \centering
    \includegraphics[width=\columnwidth]{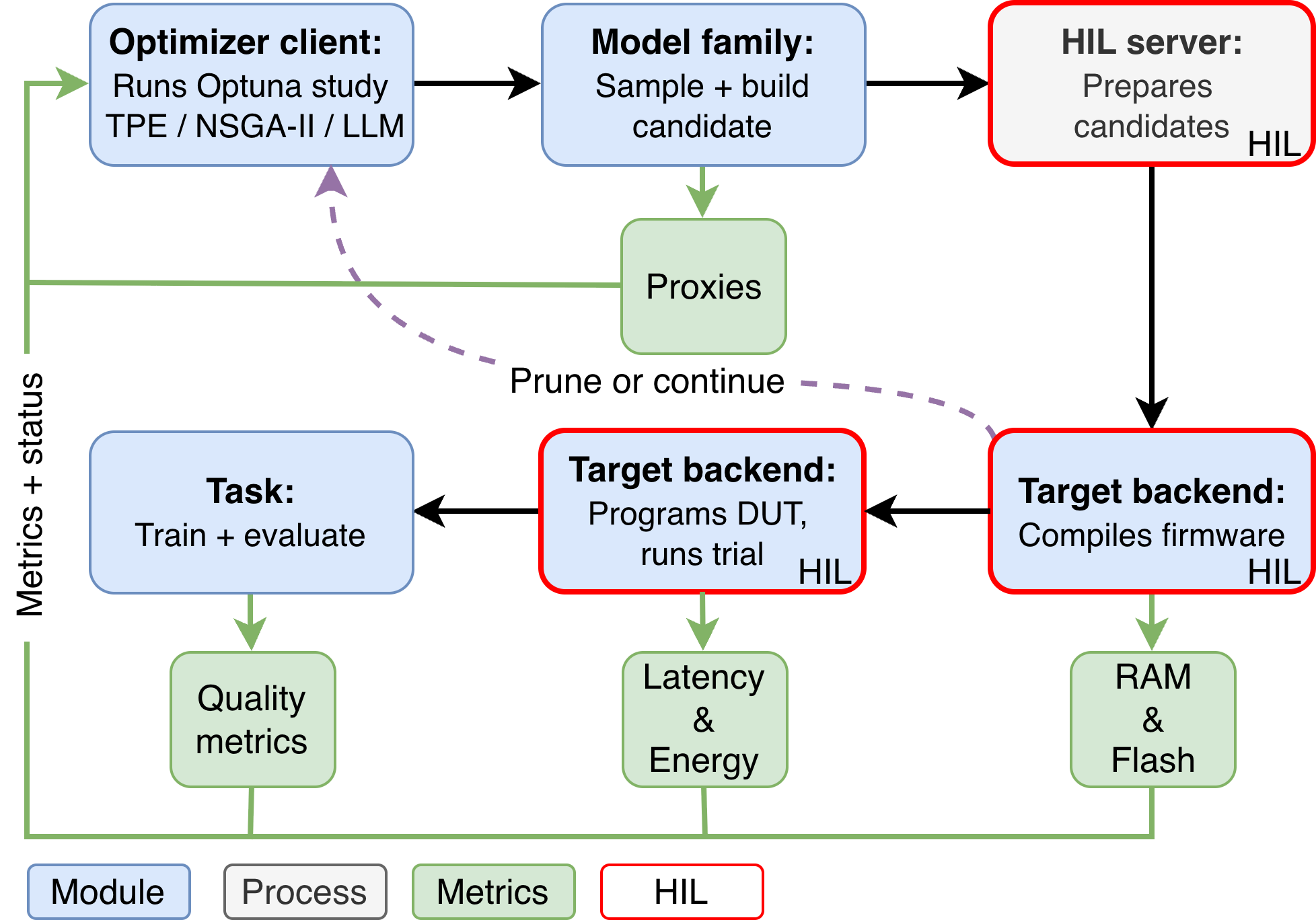}
    \caption{Per-trial implementation path in \systemname{}. Native and LLM proposals share the same evaluation path, combining host-side metrics with hardware measurements for scoring, feasibility checks and proposal feedback.}
    \Description{A flow diagram showing an optimizer client running an Optuna study with candidate proposals from TPE, NSGA-II, or an LLM. The model family samples parameters and builds a candidate. The HIL server prepares candidates, and target backends compile firmware, program the DUT, and run the trial. Quality, proxy, memory, latency, energy, and status metrics return to the client through a shared feedback path. A dashed path shows pruning or continuing based on available metrics.}
    \vspace{-1.5em}
    \label{fig:crest-nas-flow}
\end{figure}

\systemname{} evaluates candidates from native samplers and LLM proposers through the same workflow. The configured policy specifies the metrics needed for scoring, pruning and feasibility. When HIL is disabled, these requirements determine whether host evaluation is sufficient or firmware compilation is also needed. Metrics that require execution on the target require HIL. At startup, \systemname{} checks that the policy names known metrics and that the configured evaluation stages can provide them. The backend reports which measurements are available for each evaluated candidate.

\subsection{Search Engine and Trial Orchestration}
\label{subsec:impl-search-failure}

We currently support Optuna-native sampling and LLM-based proposals. Native sampling uses TPE~\cite{NIPS2011_86e8f7ab} for single-objective studies or NSGA-II~\cite{deb_fast_2002} for multi-objective studies. LLM proposal supplies explicit parameter dictionaries, which the client validates against the declared search space and queues as Optuna trials. Accepted proposals then follow the same evaluation path and produce the same measurement records as native samples. Optuna~\cite{akiba_optuna_2019} manages trials and records their outcomes in both modes.

CS3 uses two feedback-guided LLM policies called \emph{Memory} and \emph{History} (\S\ref{subsec:cs3-audio}). Memory uses a summarization scheme inspired by context compaction to retain information from older trial outcomes while keeping recent outcomes directly available. History instead supplies a compact record of all finished trials. Appendix~\ref{app:cs3-proposals} documents their prompts and proposal rules.

\systemname{} records measured candidates, policy-rejected candidates, and failures instead of dropping them from the logs. User-defined pruning and feasibility rules decide when a candidate is rejected. A candidate can also fail before measurement if it cannot compile, flash, allocate memory, or complete the runtime protocol. Transport, device-discovery, and backend setup failures can still abort a study because the measurement path itself is unavailable. In scalar studies, pruned candidates remain visible as pruned Optuna trials. In multi-objective studies, where Optuna does not support pruning, rejected candidates can receive direction-aware penalty values while the rejection reason remains in the trial log.

A separate profiling experiment evaluates ten OxIOD configurations per target, selected at evenly spaced positions along the measured-energy Pareto fronts for STM32 N657 and Portenta H7 CM7. Across 80 evaluations of these fixed configurations, continuous or 200~ms periodic HIL adds 1.3--3.3~min of preparation, deployment and measurement to 5.6--5.8~min of shared host work per candidate (Appendix~\ref{app:time-study}).

\subsection{HIL Server and Candidate Preparation}
\label{subsec:impl-hil-server}

The HIL server runs on the machine connected to the DUT and receives each candidate from the optimizer client through a ZeroMQ request--reply connection. The client sends the candidate hyperparameters, runtime metadata, quantization mode and target options. The server builds the export model, checks its input shape and task output, and sends it to the selected backend. The optimizer can run on the same host or on a separate GPU server.

Each trial resolves one model architecture, quantization mode, and export variant before host TFLite evaluation or backend preparation. This keeps the host-evaluated and deployed candidates aligned because they share the same model architecture and quantization mode, while the export variant determines the concrete model artifact sent to the backend. Backends then perform device-specific conversion, staging, build, upload, and measurement.

\subsection{HIL Harness and Power Instrumentation}
\label{subsec:impl-hil-harness}

\begin{figure}[t]
    \centering
    \includegraphics[width=\columnwidth]{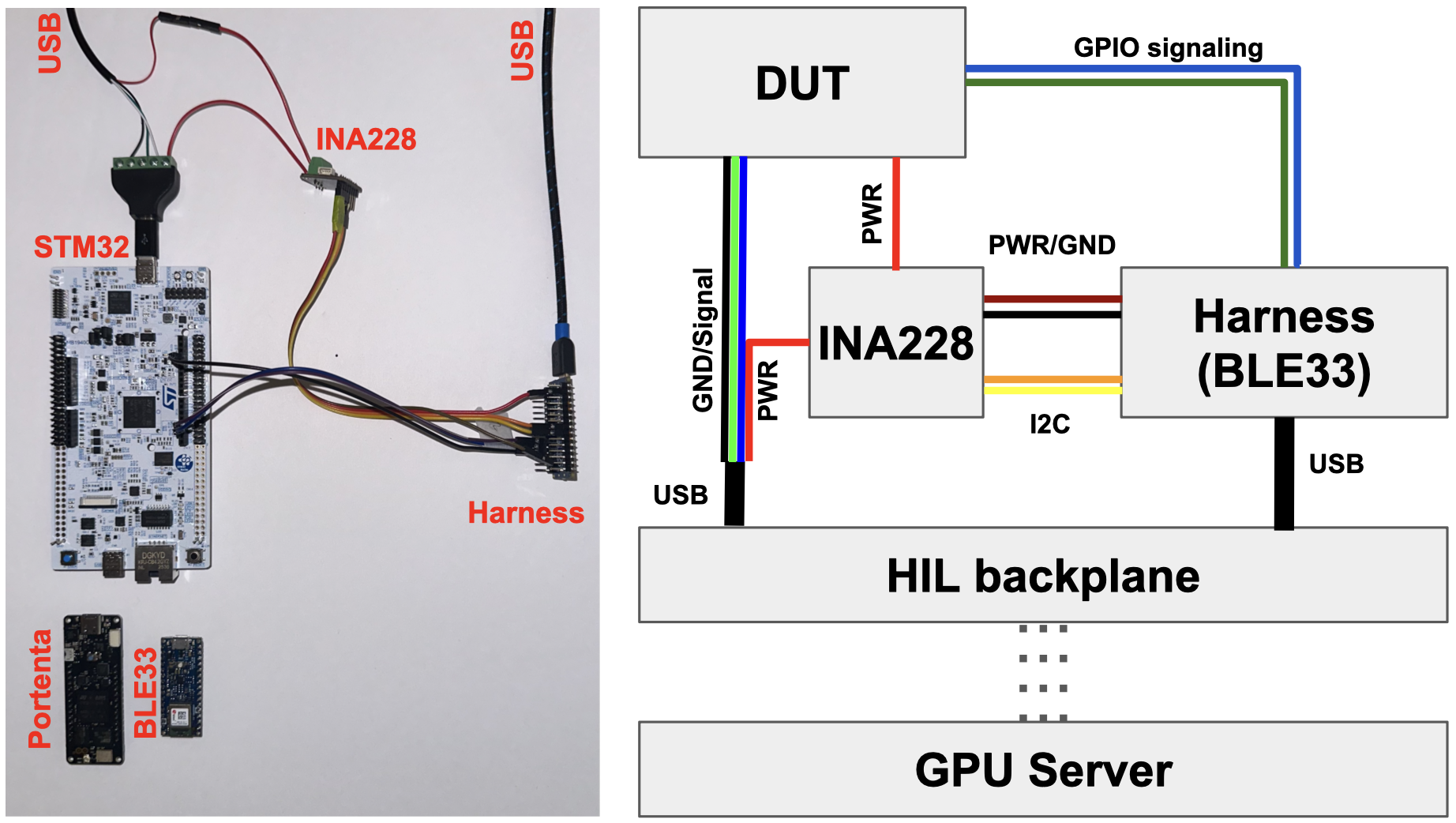}
    \caption{\systemname{} HIL power-measurement setup. The DUT is powered through an INA228 monitor, and a separate harness MCU reads power telemetry over I2C while observing the DUT's GPIO markers.}
    \Description{A diagram showing a DUT powered through an INA228 sensor, a BLE33 harness connected to the INA228 over I2C, GPIO signaling wires from the DUT to the harness, USB connections from the DUT and harness to a HIL backplane, and a GPU server connected below the backplane.}
    \vspace{-1.5em}
    \label{fig:hil-harness-backplane}
\end{figure}

Figure~\ref{fig:hil-harness-backplane} shows the HIL power-measurement setup. \systemname{} adapts Bartoli et al.'s trigger-delimited TinyML energy-measurement method to a per-trial HIL NAS harness~\cite{bartoli_benchmarking_2025}. Their setup uses an oscilloscope to observe shunt voltage and DUT trigger signals. \systemname{} instead powers the DUT through an off-the-shelf Adafruit INA228 breakout board~\cite{industries_adafruit_webpage}. A separate Arduino Nano 33 BLE Sense harness MCU reads the INA228 over I2C using the Adafruit INA228 Arduino library~\cite{adafruitadafruit_ina228_library} and observes DUT GPIO markers.

The GPIO markers define the measurement window. This lets \systemname{} attach each energy value to a specific candidate, runtime schedule, and trial outcome without manual trace segmentation or signal processing to infer where inference occurred. Continuous-inference runs mark the repeated inference interval. Periodic runs mark the full measurement window, including inference and the sleep or waiting intervals.

The harness provides an accessible, low-cost measurement path using commercially available components. It measures the evaluated development boards through their USB supply. Readings include all components powered through that input, including the STM32 board's ST-LINK debugger. At this supply boundary, the evaluated development boards retain substantial power draw during sleep. Consequently, replacing inference time with waiting can save relatively little energy, making the waiting policy and measurement boundary important to configuration selection. Appendix~\ref{app:measurement-context} reports whole-board energy probes in sleep and active states. The same evaluation workflow can instrument an accessible MCU rail or custom-board supply, with appropriate power routing and backend integration. Energy values and rankings must then be measured for that deployment's power boundary and runtime policy.

The harness uses a 15~m\(\Omega\) shunt, 0.30~A expected-current range, 40.96~mV ADC range, 280~\(\mu\)s current conversion, 150~\(\mu\)s voltage conversion, and 64-sample averaging~\cite{adafruitadafruit_ina228_library,texas_instruments_ina228_nodate}. At the GPIO-marked start of each measurement window, the harness resets the INA228 energy accumulator, and at the end marker it reads the accumulated energy value and normalizes it by the completed invocation count. The INA228 accumulates energy on each conversion cycle independently of the configured sample averaging~\cite{texas_instruments_ina228_nodate}. On coordinated DUT-plus-harness paths, harness energy is accepted only when DUT timing exists, harness completion is observed, and DUT and harness run counts match. Harness-only paths rely on harness completion and harness telemetry.

\subsection{Backend Implementations}
\label{subsec:impl-backends}

\systemname{} provides backends for the Arduino Nano 33 BLE Sense (Cortex-M4)~\cite{arduino_nano33ble_sense_datasheet}, Arduino Portenta H7 (Cortex-M7 and Cortex-M4)~\cite{arduino_portenta_h7_datasheet}, and STM32 NUCLEO-N657X0-Q (Cortex-M55)~\cite{st_stm32n6_datasheet}. Each backend builds and programs candidates, collects measurements and status, and handles cleanup for its target.

The Arduino targets share an \texttt{arduino-cli} build and upload path. Portenta H7 also requires target-core handling because its two cores use different execution and measurement paths. STM32 N657 uses ST Edge AI and the ST toolchain, with different memory reports, programming tools and telemetry. The backends return these results through common metric fields used by search, policy evaluation and replay.

The STM32 backend supports CPU-clock selection and exposes clock, schedule and sleep-state telemetry. The Arduino backends do not expose CPU-clock selection and provide fewer runtime fields. Measurement checks reflect these differences. On coordinated Arduino DUT-plus-harness paths, energy is accepted only when DUT timing exists and DUT and harness run counts match. STM32 additionally checks the requested phase, configured clock, RTC window, STOP-entry counts and invocation count. An instrumentation failure can trigger one retry using the compiled candidate, with both attempt records retained. A valid measurement that exceeds a latency limit or misses a release remains available for analysis, while the configured feasibility rules determine whether the candidate can be selected.

The backends also implement the continuous and periodic schedules defined in \S\ref{subsec:runtime-schedules}. Execution and measurement use direct serial control, coordinated DUT-plus-harness telemetry or harness-only telemetry, depending on the target. The evaluation reports only the fields supported by each measurement path.

For periodic execution, the Arduino targets use light sleep to preserve the USB connection for trial coordination and the \texttt{micros()} timebase for release scheduling. The Portenta H7 deeper-sleep library disables USB and stops that timer~\cite{noauthor_arduino_lowpowerportentah7_nodate}.

The STM32 N657 backend uses system STOP mode through the wait-for-interrupt instruction (WFI), with wake-up scheduled by the real-time clock (RTC)~\cite{noauthor_stm32n6_2025}. The RTC uses the board's 32.768~kHz low-speed external crystal independently of the CPU clock. Periodic measurements span ten release periods. When slack exceeds 10~ms, the runtime enters STOP and wakes about 5~ms before the next release, then polls until release. Appendix~\ref{app:cs1-waiting-model} describes the waiting policies used in each experiment.

\subsection{Logging and Replay}
\label{subsec:impl-logging-replay}

\systemname{} records the selected configuration, candidate payloads, per-trial CSV logs, and selected artifacts. Trial logs include task metrics, hyperparameters, objective metadata, feasibility metadata, quantization mode, hardware metrics, status fields, and periodic telemetry when present. Together these fields store the measurement record for each trial (\S\ref{subsec:measurement-records}). Multi-objective studies export Pareto-front candidates for analysis and replay, while scalar studies track the best feasible candidates under the score. Replay reconstructs candidate requests and reuses candidate preparation and backend evaluation without running the proposer or repeating training. Its records associate inherited quality with the source evaluation and new hardware costs with the destination target. CS3 uses this replay path to evaluate selected configurations on another target.

\subsection{Use of AI Tools}
\label{subsec:ai-tools}

We used GitHub Copilot, OpenAI Codex and Anthropic Claude Code to assist with implementation, testing, analysis scripts and plotting. This included the HIL server, target backends, DUT firmware, unit tests and CS3 proposal prompts and integration code. All energy, latency, and quality values come from hardware measurements and model training. The authors designed the system and experiments, reviewed the resulting code, analyses and text, and take responsibility for the final work. The LLM candidate proposal policies used in CS3 are described in \S\ref{subsec:cs3-audio} and Appendix~\ref{app:cs3-proposals}.

\section{Evaluation}
\label{sec:evaluation}

We evaluate \systemname{} through three case studies on two sensing workloads across three Arm Cortex-M development boards. CS1 investigates how runtime schedules affect model and clock selection and evaluates calibrated energy estimates. CS2 and CS3 examine target-dependent feasibility and energy tradeoffs while reusing the evaluator across workloads and candidate proposers.

The first workload is inertial odometry using the Oxford Inertial Odometry Dataset (OxIOD)~\cite{chen_oxiod_2018}, with a temporal convolutional network family similar to prior neural inertial odometry work~\cite{saha_tinyodom_2022}. The second is audio classification on UrbanSound8K~\cite{salamon_dataset_2014}, using a depthwise-separable CNN family in the lineage of compact MCU audio models~\cite{zhang_hello_2018,nordby_environmental_2019,bartoli_end--end_2025}. We evaluate these workloads on the Arduino Nano 33 BLE Sense, Arduino Portenta H7 and STM32 N657, reporting the Portenta CM7 and CM4 cores as separate target configurations where evaluated. We use recorded datasets to train candidates and evaluate prediction quality, and measure deployment costs on the target boards through the shared evaluation path described in \S\ref{sec:system} and \S\ref{sec:technical-implementation}. Measurements cover inference on recorded inputs or prepared features and, in periodic mode, the following waiting intervals. Sensor acquisition and feature extraction lie outside this measurement boundary.

We cap candidate training at 55 epochs to bound search cost. For 50 selected OxIOD configurations from the CS2 Portenta CM7 search, validation-error rankings at this budget had a Spearman correlation of 0.842 with rankings at a 300-epoch reference budget. Appendix~\ref{app:training-budget} examines the training-cost and selection tradeoff. Optuna~\cite{akiba_optuna_2019} manages trials, with budgets reported per study.

For OxIOD, the task quality metric is aggregate odometry error,
\begin{equation}
  \mathrm{RMSE}_{\mathrm{total}} =
  \sqrt{\frac{1}{n}\sum_{i=1}^{n}(\hat{v}_{x,i} - v_{x,i})^2}
  +
  \sqrt{\frac{1}{n}\sum_{i=1}^{n}(\hat{v}_{y,i} - v_{y,i})^2}.
  \label{eq:oxiod-rmse}
\end{equation}
Here, $n$ is the number of evaluated samples, and $\hat{v}_{x,i}$ and $\hat{v}_{y,i}$ are predicted velocity components with reference values $v_{x,i}$ and $v_{y,i}$. Audio classification uses validation macro-F1.

Search measurements use windows of ten invocations in continuous mode or ten releases in periodic mode. Continuous measurements report energy per inference and latency, while periodic measurements report energy per release and available timing and deadline telemetry. Other probe protocols are described with their experiments. Appendix~\ref{app:trial-outcomes} summarizes the recorded trial outcomes.

\subsection{Case Study 1. Inference Savings Do Not Always Transfer to Periodic Execution}
\label{subsec:cs1-schedule}

Inference savings do not always transfer to periodic execution. Waiting can absorb most of the advantage, leaving small energy differences that decide the preferred configuration. Some of these differences lie outside the evaluated calibrated estimates. CS1 examines this effect on STM32 N657. We compare continuous energy and latency, objectives used by prior HIL search~\cite{groh_edgevolution_2026,saha_tinyodom_2022,saha_tinyns_2024}, with measured energy per release and the estimates in \S\ref{sec:energy}.

We run six searches, with a continuous-energy search and a periodic-energy search at each release period of 100, 200 and 500~ms. Each search also optimizes validation RMSE, and we analyze its first 256 trial attempts. Controlled replay then compares configurations from these searches under different clocks and waiting policies.

\textbf{Waiting offsets inference savings.} For the fixed int8 TCN architecture from trial 203 of the 500~ms continuous-energy search, the fitted decomposition assigns 8.80~mJ to additional waiting at 800~MHz, offsetting 97\% of its continuous-energy advantage over 300~MHz. This follows the accounting in Equation~\ref{eq:crest-release-accounting}. The faster configuration finishes in about 6.70 rather than 17.70~ms, leaving 11~ms more waiting time.

Table~\ref{tab:metadata-interval-example} shows how the choice depends on the schedule. Continuous execution favors trial 203 at 800~MHz, a 100~ms schedule favors the same model at 300~MHz, and a 500~ms schedule favors trial 137 at 300~MHz. Both architectures come from the 500~ms continuous-energy search, and all three configurations satisfy a validation RMSE ceiling of 1.0. The comparisons use STOP in every waiting interval. Appendix~\ref{app:cs1-waiting-model} describes the measurement procedure.

\begin{table}[b]
\centering
\vspace{-0.4em}
\caption{Three configurations from the direct policy experiment with STOP between releases. Trial numbers refer to the 500~ms continuous-energy search. Values are means over five paired rounds. Bold marks the lowest energy in each column.}
\label{tab:metadata-interval-example}
\vspace{-1em}
\small
\setlength{\tabcolsep}{3pt}
\begin{tabular*}{\columnwidth}{@{\extracolsep{\fill}} l r r r r @{}}
\toprule
Configuration & RMSE & \makecell{Continuous\\(mJ/inference)} & \makecell{100~ms\\(mJ/release)} & \makecell{500~ms\\(mJ/release)} \\
\midrule
Trial 203, 800~MHz & 0.660 & \textbf{8.01} & 83.58 & 405.10 \\
Trial 203, 300~MHz & 0.660 & 17.06 & \textbf{83.32} & 403.52 \\
Trial 137, 300~MHz & 0.912 & 47.36 & 85.45 & \textbf{401.85} \\
\bottomrule
\end{tabular*}
\end{table}

Continuous energy favors 800~MHz, using 8.01 rather than 17.06~mJ per inference. Energy per release instead favors 300~MHz by 0.26~mJ at 100~ms and 1.58~mJ at 500~ms, with paired 95\% intervals of 0.13--0.38 and 0.77--2.40~mJ. Because waiting offsets most of the inference saving, the preferred clock is decided by margins of approximately 0.3\% and 0.4\% of release energy. Although these savings are small, they reverse the clock choice made from continuous energy under both release periods.

\textbf{Shared waiting parameters miss a change in architecture choice.} A shared waiting model predicts that changing the period preserves the preferred architecture at a fixed clock. In Equation~\ref{eq:crest-window-energy-current}, both candidates receive the same $T\widehat P_{\mathrm{wait}}(f)$ contribution. With the remaining terms fixed, changing $T$ therefore preserves their ordering. The measurements instead favor trial 203 at 100~ms and trial 137 at 500~ms. At 500~ms, trial 137 saves 1.67~mJ per release, with a paired 95\% interval of 0.91--2.43~mJ. Their fitted period slopes differ by 9.43~mW, with a paired 95\% interval of 7.47--11.39~mW. This comparison is paired within one measurement session. Appendix~\ref{app:cs1-waiting-model} gives the fitted slopes and corrections.

\begin{figure}[t]
\centering
\includegraphics[width=0.8\columnwidth]{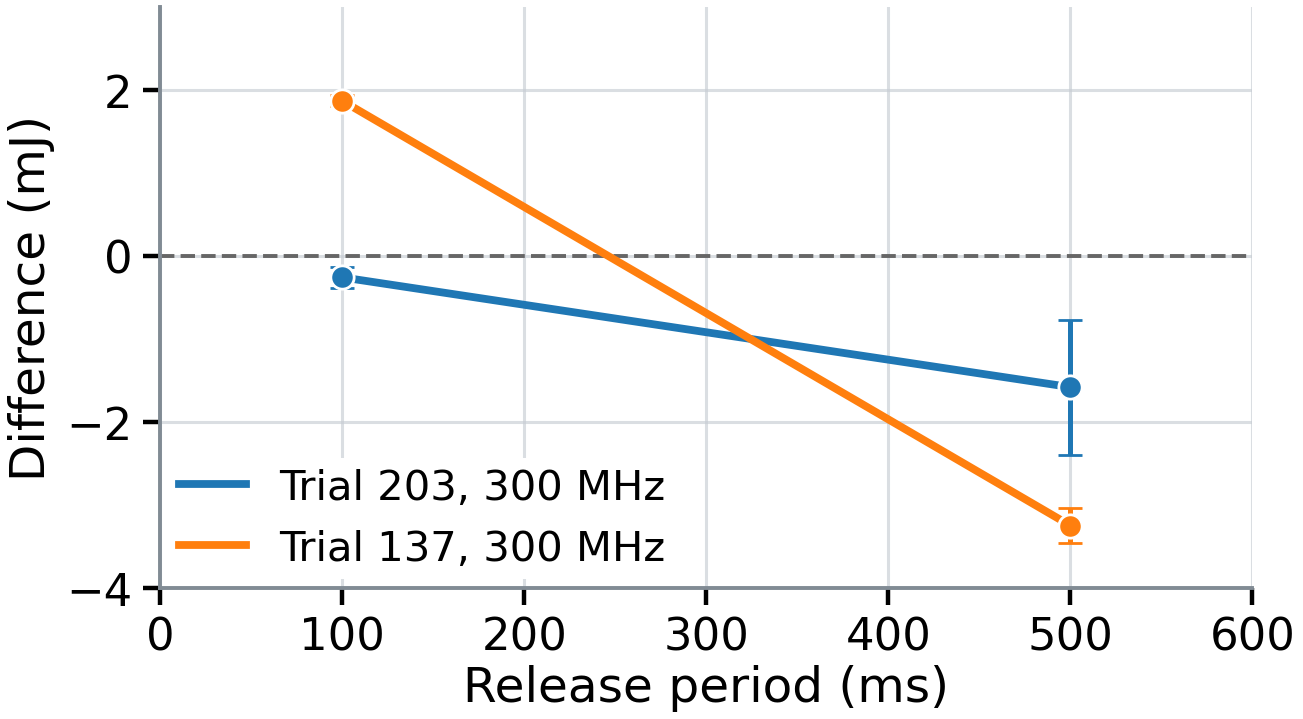}
\vspace{-0.5em}
\caption{Energy per release relative to trial 203 at 800~MHz. Negative values favor 300~MHz. Points show means and paired 95\% intervals over five rounds with STOP between releases. Lines connect measured endpoints.}
\Description{Two lines show release-energy differences at 100 and 500 ms relative to trial 203 at 800 MHz. The measured endpoints show a change in the preferred architecture at 300 MHz.}
\vspace{-1.5em}
\label{fig:cs1-clock-reversal}
\end{figure}

Figure~\ref{fig:cs1-clock-reversal} plots both 300~MHz configurations relative to the continuous-energy choice at 800~MHz. The lower line identifies the better 300~MHz configuration at each period. Their crossing shows the change in architecture choice.

\textbf{Release-energy error remains with measured continuous energy.} Across the search pools, latency predicts continuous energy well, but substantial release-energy error remains even when continuous energy is measured.

\begin{figure}[t]
\centering
\includegraphics[width=0.8\columnwidth]{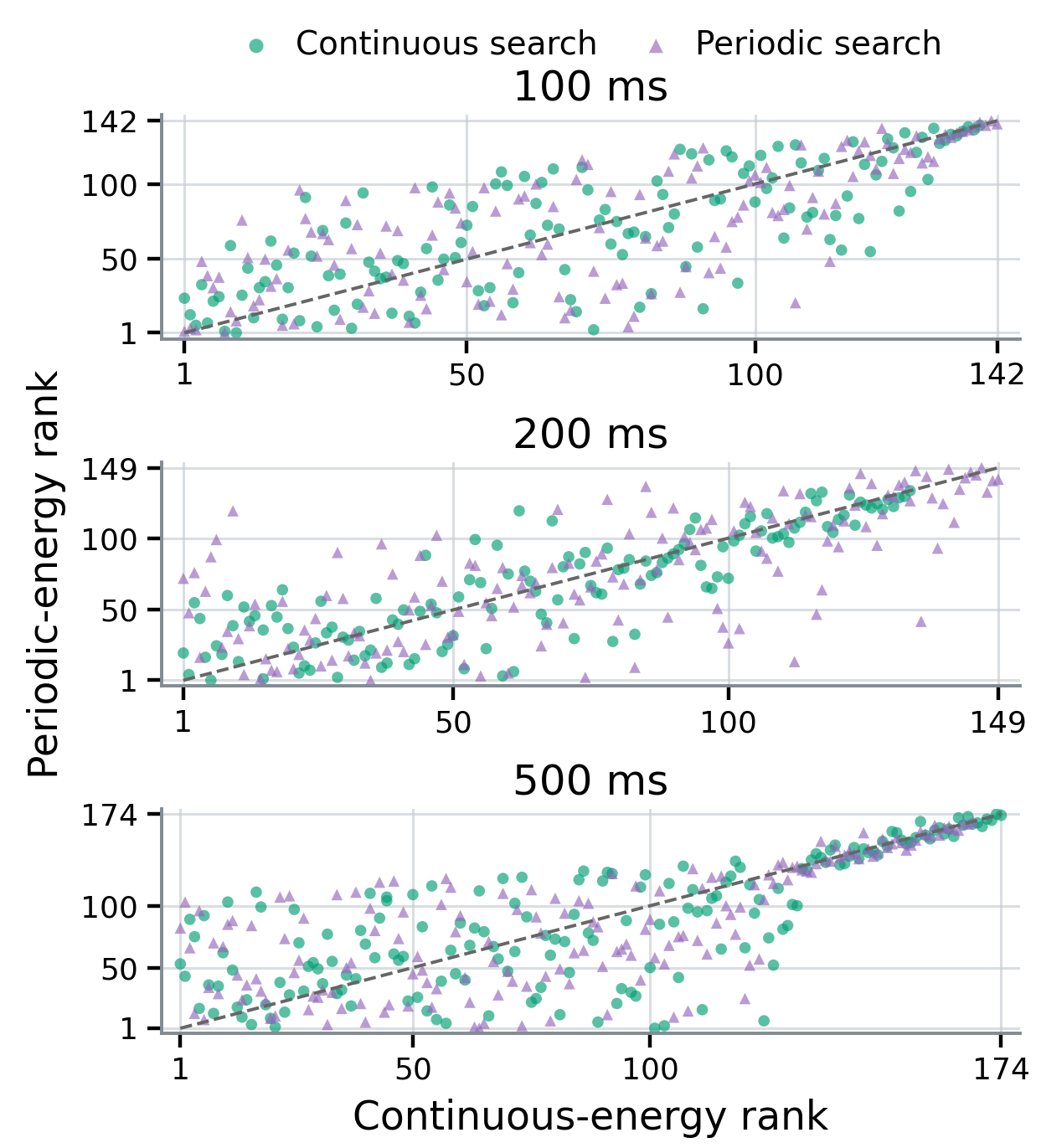}
\vspace{-0.5em}
\caption{Energy ranking changes across search candidates. Rank 1 is lowest energy. Points on the diagonal keep their rank. Both continuous and periodic search pools are shown at each period.}
\Description{Three panels plot continuous-energy rank against periodic-energy rank at 100, 200 and 500 ms. Green circles show continuous-search observations and purple triangles show periodic-search observations at every period, ranked separately within each pool. Points away from the diagonal change rank.}
\vspace{-1.5em}
\label{fig:cs1-energy-ranks}
\end{figure}

Figure~\ref{fig:cs1-energy-ranks} shows 905 paired observations with valid timing and no deadline misses across the six search pools. Continuous and periodic energy give different rankings in both search types. The search windows leave the final waiting interval awake.

The calibration comparison follows the predictors in \S\ref{sec:energy}. Equation~\ref{eq:crest-window-energy-current} uses measured continuous energy through $P_{\mathrm{cont}}(x)C(x)$. The latency-based alternative replaces that term with the affine estimate $a_{f,q}+s_{f,q}C(x)$ from Equation~\ref{eq:crest-affine-continuous}. A proportional baseline sets the intercept to zero. All three use the same probe-calibrated waiting power $\widehat P_{\mathrm{wait}}(f)$ and correction $\widehat b(f)$. A fourth estimate fits release energy directly from latency using scheduled calibration measurements at the same period and policy.

We fit the latency-based estimates separately within each pool, frequency and quantization group, using architectures excluded from evaluation. Repeated five-fold splits keep all copies of an inference architecture together. Table~\ref{tab:cs1-heldout-prediction} reports root mean squared (RMS) error against the held-out measurements. Its continuous column evaluates the latency-based inference-energy fits, while its per-release column evaluates all four estimates of scheduled energy. The measured-continuous row reports only release-energy error because it uses measured continuous energy as an input. Comparing this row with the fitted rows shows how much release-energy error remains when the continuous term is measured rather than predicted. Appendix~\ref{app:cs1-heldout} gives the calibration support, weighting and split details.

\begin{table}[b]
\centering
\vspace{-0.4em}
\caption{Root mean squared prediction error over held-out predictions, in mJ. Lower values indicate better prediction accuracy. Continuous-derived estimates use the same waiting calibration. The direct scheduled affine estimate uses scheduled-energy calibration labels from the same pool.}
\label{tab:cs1-heldout-prediction}
\vspace{-1em}
\small
\setlength{\tabcolsep}{3pt}
\begin{tabular*}{\columnwidth}{@{\extracolsep{\fill}} l r r @{}}
\toprule
Estimate & Continuous & Per release \\
\midrule
Fitted proportional & 1.57 & 3.99 \\
Fitted affine & 1.38 & 3.92 \\
Measured continuous & -- & 3.48 \\
Direct scheduled affine & -- & 3.11 \\
\bottomrule
\end{tabular*}
\end{table}

The affine continuous estimate has a median group RMS error of 1.69\% of group mean energy. Measuring continuous energy instead of estimating it from latency reduces release-energy RMS error from 3.92 to 3.48~mJ. Direct scheduled calibration gives the lowest overall release-energy RMS error among these estimates. The remaining 3.48~mJ RMS error with measured continuous energy in Equation~\ref{eq:crest-window-energy-current} is larger than the margins that decided the clock choice above. These residuals can include measurement variation and scheduled execution costs that the shared waiting power and correction do not capture. Paired replay below tests whether particular differences persist across rounds.

We also compare their selections under validation RMSE ceilings of 0.65, 0.75 and 1.0, measuring regret against the lowest measured energy per release among eligible candidates. Selecting by continuous energy alone gives mean observed regret of 2.64~mJ, and selecting by latency alone gives 2.77~mJ. Adding the waiting term lowers mean regret by about 1~mJ, to 1.72~mJ with the continuous affine estimate and 1.64~mJ with measured continuous energy. The direct scheduled estimate gives 1.73~mJ. The affine and measured-continuous estimates select equal measured energy in 542 of 622 comparisons. Thus, estimating the continuous term from latency often preserves the selection obtained by measuring that term. Direct scheduled calibration lowers prediction error without lowering mean regret, so prediction accuracy and selection quality are distinct targets. Appendix~\ref{app:cs1-selection} reports the comparison.

\textbf{Repeated measurements reveal differences missed by the estimates.} The question is whether candidates with similar continuous latency differ in scheduled execution in ways that the shared-parameter estimates miss. Paired replay tests whether these differences persist across repeated measurements. Six pairs were randomly selected using search-time latency without selecting on energy gaps. One of the six, trials 228 and 243 from the 500~ms periodic-energy search at 400~MHz, showed a resolved reversal between continuous and release-energy ordering. Appendix~\ref{app:cs1-transfer-replay} reports all six pairs and their prediction residuals.

We compare this pair and trials 110 and 123 from the same 500~ms periodic-energy search under the awake-final and STOP-final protocols in ten paired rounds at 500~ms. Trial 228 uses 0.57~mJ more continuous energy than 243, but 3.73~mJ less per release under the awake-final protocol and 5.81~mJ less under the STOP-final protocol. The STOP-final advantage has a paired 95\% interval of 5.71--5.92~mJ. The 228/243 reversal retains its direction in every round under both protocols. The other pair's release-energy gap shrinks under STOP-final execution. Trial 228 does not meet the CS1 validation RMSE ceilings, so this pair is an energy-ordering counterexample rather than a selection change. Its 5.81~mJ difference exceeds the 0.26--1.58~mJ margins that decided the clock choice above, showing that scheduled differences missed by the estimates can be as large as the margins that decide selection. Appendix~\ref{app:cs1-final-policy-replay} reports both pairs and their timing.

\textbf{\systemname{} impact.} These counterexamples establish limits of the evaluated predictors; scheduled measurement provides a way to check deployment choices without assuming those predictors capture every relevant cost. \systemname{}'s modular design lets these studies vary the selection objective and runtime schedule while retaining the workload and model family and reusing candidate preparation, measurement and replay.

\begin{figure*}[t]
\centering
\includegraphics[width=\textwidth]{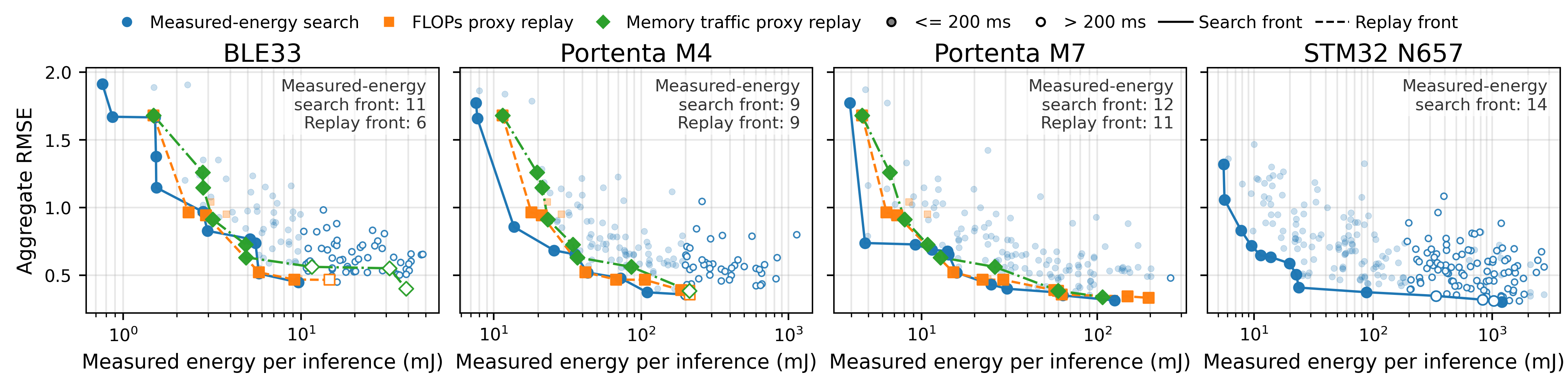}
\vspace{-0.5em}
\caption{Proxy Pareto candidates replayed on the three Arduino target configurations and measured-energy search fronts on all four targets. Orange dashed curves show FLOPs selections, green dash-dot curves show memory-traffic selections, and solid blue curves show measured-energy selections. The STM32 N657 panel shows the measured-energy search over model and CPU-clock choices. Energy per inference uses a log scale. Lower energy and RMSE are better. Filled markers meet the 200~ms latency threshold. Replay-front counts in each panel refer to the FLOPs front.}
\Description{Four side-by-side scatter plots. The first three compare measured-energy NAS fronts with FLOPs-proxy and memory-traffic-proxy replay fronts for BLE33, Portenta M4, and Portenta M7. In each of these panels, the measured-energy front lies at lower energy, lower RMSE, or both for many replayed proxy-front points. The fourth panel shows the STM32 N657 measured-energy search front over model and CPU-clock choices.}
\vspace{-1.0em}
\label{fig:cs2-proxy-replay}
\end{figure*}

\subsection{Case Study 2. Target Measurements Change Proxy-Selected Tradeoffs}
\label{subsec:cs2-proxy}

Static proxy objectives identify candidate tradeoffs, but target deployment determines which candidates fit and which remain competitive in measured energy. CS2 tests this distinction by replaying proxy-selected candidates and comparing their measured fronts with candidates from measured-energy search. We hold the OxIOD workload and TCN family fixed and compare FLOPs, memory-traffic and measured-energy objectives on BLE33 and the Portenta H7 M4 and M7 cores. We then extend the measured-energy search to STM32 N657 with CPU clock as an additional candidate parameter.

\textbf{Comparison setup.} All searches optimize validation RMSE alongside their cost objective using Optuna's NSGA-II sampler~\cite{akiba_optuna_2019,deb_fast_2002}. Proxy and measured searches use the same architecture and float or int8 search space. The TCN family spans 8.1M hyperparameter combinations across filters, kernel sizes, dropout rates, skip connections, normalization and dilation patterns. We run one search per objective and target setting. The proxy and Arduino HIL searches each have a budget of up to 150 attempts. This budget bounds experiment time for the proxy comparison and is a resource allocation rather than a convergence-based stopping criterion. Appendix~\ref{app:trial-outcomes} reports completed and recorded attempts.

Proxy searches compute static costs without target compilation or explicit capacity-pruning rules. Replay then checks deployment through the target backend, including flash, RAM and tensor-arena limits. The 200~ms latency threshold corresponds to OxIOD's configured stride of 20 samples at 100~Hz. We report this threshold alongside the continuous-energy comparisons rather than using it to filter their Pareto fronts.

\textbf{Target deployment changes the eligible set.} Each proxy front contributes 16 unique candidates, and each set is replayed on every target. The FLOPs selection contains 12 float and four int8 candidates, while memory-traffic selection contains five float and 11 int8 candidates. Replaying each set on three Arduino target configurations gives 48 target evaluations, with 35 valid measurements for FLOPs and 36 for memory traffic. Each failed candidate exhausted the tensor arena or exceeded flash. On BLE33, only nine of the 16 FLOPs-selected candidates deploy successfully. Recomputing their measured Pareto front leaves six points.

The FLOPs objective uses TensorFlow's operation count for the frozen batch-size-one graph, without target-specific operator costs. The memory-traffic objective uses the static byte count $T(m,q)$. For each Keras layer $\ell$, let $I_\ell$, $O_\ell$ and $W_\ell$ denote the scalar counts of its batch-size-one input activations, output activations and unique weights. With scalar byte width $b(q)$ under deployment precision $q$, the proxy is
\begin{equation}
  T(m,q)=b(q)\sum\nolimits_{\ell\in m}(I_\ell+O_\ell+W_\ell).
  \label{eq:memory-traffic-proxy}
\end{equation}
This count includes graph-level tensor traffic but does not model cache behavior, kernel scheduling or backend-specific memory placement~\cite{ma_shufflenet_2018,lai_not_2018,lin_mcunet_2020}. FLOPs does not distinguish deployment precision, while the memory-traffic proxy scales by byte width.

TinyOdom informs our TCN family and uses Mango to optimize a scalar combination of validation error, compiler-reported memory and measured latency~\cite{saha_tinyodom_2022}. \systemname{} supports scalar and Pareto objectives, including measured energy per release. CS1 compares latency and energy objectives under common quality constraints. This isolates cost-signal differences without reproducing TinyOdom's complete search pipeline. CS2 compares proxy and measured-energy search through the same evaluator.

\textbf{Target measurements change the available tradeoffs.} Among candidates that deploy, the relevant comparison is their measured energy at a given prediction quality. We recompute each target's proxy-replay front using valid measured energy and inherited source validation RMSE, then compare it with the measured-search front. A replayed point is dominated if a measured-search candidate has no higher energy or RMSE and is strictly better in at least one.

For each point on a target's replay front, we select the measured-search Pareto point with the closest RMSE. The percentage energy gap is $100(E_{\mathrm{proxy}}-E_{\mathrm{measured}})/E_{\mathrm{proxy}}$, with positive values favoring measured search. Median absolute RMSE mismatches are 0.013--0.018 for FLOPs and 0.031--0.045 for memory traffic. Table~\ref{tab:cs2-results} reports the median gap among valid replay-front points for each target. We summarize across targets using the median of these three values.

\begin{table}[b]
\centering
\vspace{-1em}
\caption{Arduino proxy-replay results. Valid replay counts are out of the 16 unique source candidates for each proxy. Dominated and $\le$200 ms counts are out of valid replay-front points for each target configuration.}
\label{tab:cs2-results}
\vspace{-0.5em}
\footnotesize
\setlength{\tabcolsep}{3pt}
\begin{tabular*}{\columnwidth}{@{\extracolsep{\fill}} l l c c c c @{}}
\toprule
Proxy & Board & \makecell{Valid\\replays} & Dominated & $\le$200 ms & Energy gap \\
\midrule
FLOPs & BLE33 & 9/16 & 3/6 & 5/6 & 2.1\% \\
FLOPs & Portenta M4 & 12/16 & 5/9 & 7/9 & 31.5\% \\
FLOPs & Portenta M7 & 14/16 & 5/11 & 11/11 & 14.8\% \\
Mem. traffic & BLE33 & 11/16 & 6/9 & 6/9 & 45.2\% \\
Mem. traffic & Portenta M4 & 12/16 & 7/8 & 7/8 & 38.4\% \\
Mem. traffic & Portenta M7 & 13/16 & 3/8 & 8/8 & 27.1\% \\
\bottomrule
\end{tabular*}
\end{table}

The first three panels of Figure~\ref{fig:cs2-proxy-replay} place the proxy-replay fronts alongside the measured-search fronts for each target. Their overlap and separation show where the objectives produce similar or different measured tradeoffs. Measured-energy candidates dominate 13 of the 26 FLOPs replay-front points. Of those 26 replay-front points, 23 meet the 200~ms latency threshold. The median nearest-RMSE energy gap across targets is 14.8\%. For memory traffic, measured-energy candidates dominate 16 of the 25 replay-front points, 21 replay-front points meet the threshold, and the median gap across targets is 38.4\%.

\textbf{The value of a proxy depends on the target.} FLOPs selection has a median nearest-RMSE energy gap of 2.1\% on BLE33 and 31.5\% on Portenta M4. The fronts also contain complementary tradeoffs. FLOPs replay dominates one measured-front point on BLE33 and one on Portenta M4. Memory-traffic replay dominates two on BLE33 and two on Portenta M7. These reverse-dominance counts are unchanged when both candidates must meet the 200~ms threshold. The measured-search and proxy-replay fronts therefore provide different candidate tradeoffs, whose energy and feasibility must be established on the intended target.

\textbf{The same evaluator supports target-specific choices.} The fourth panel reports a separate STM32 N657 search over architecture, precision and CPU clock, outside the Arduino proxy comparison. Its 250-attempt budget covers clock choices of 200, 300, 400, 600 and 800~MHz. The search produces 242 valid measured observations and a 14-point energy--RMSE front, shown in the fourth panel of Figure~\ref{fig:cs2-proxy-replay}. Eight of the 14 front observations use 800~MHz, compared with 50 of the 242 valid observations. CS1 shows how including waiting energy can change this clock choice.

\textbf{\systemname{} impact.} A proxy front alone does not establish which tradeoffs are available on the deployment target. \systemname{} takes candidates selected under different objectives through common deployment checks and measurements, then compares their measured energy and source quality. This connects proxy selection to target feasibility and energy while retaining the same workload, candidate representation and replay workflow.

\subsection{Case Study 3. Reusing the Evaluator Across Workloads and Transferring Selections Across Targets}
\label{subsec:cs3-audio}

A common evaluator lets different proposers select candidates under the same application objective, while replay tests whether their energy ordering transfers to another target. CS3 demonstrates this workflow on UrbanSound8K audio classification with TPE and two LLM proposal strategies on STM32 N657 and Portenta H7 CM7. We replace OxIOD and TCNs with a dataset adapter for cached log-mel features, macro-F1 evaluation and a DS-CNN family through the interfaces in \S\ref{subsec:adding-workloads-models}. Within each target, all three proposers retain the existing deployment and measurement backends, application score and deployment checks.

\textbf{Setup.}
We use a fixed split for search, with folds 1 to 8 for training, fold 9 for validation and fold 10 reserved for final testing. This gives candidate selection a common validation set while preserving UrbanSound8K's predefined fold boundaries~\cite{salamon_dataset_2014}. Audio candidates all use int8 PTQ. Candidate training uses the shared 55-epoch cap.

We compare Optuna's TPE sampler~\cite{NIPS2011_86e8f7ab} with the Memory and History policies introduced in \S\ref{subsec:impl-search-failure}. Both LLM policies use OpenAI's GPT-6 Luna. Appendix~\ref{app:cs3-proposals} gives their prompts and proposal rules.

Each strategy uses the same application score
\begin{equation}
S = \mathrm{MacroF1}_{\mathrm{val}}
  - 0.10\frac{E_{\mathrm{norm}}}{400\,\mathrm{mJ}} .
\label{eq:cs3-audio-score}
\end{equation}
Here, $E_{\mathrm{norm}}$ is continuous energy per inference on the development board, normalized as in Equation~\ref{eq:crest-energy}.

An inference using 400~mJ receives a penalty of 0.10 in macro-F1 units. Latency and memory are checked separately. Appendix~\ref{app:cs3-proposals} defines this scoring scale and reports weight sensitivity.

\textbf{Search budget and cost.}
We ran each strategy once per target, for six total searches. Each search was allowed up to 300 trial attempts to obtain 200 feasible trials. A trial counts toward the 200 if it completes without pruning, produces a score, and passes the 2~s latency limit and the target's RAM and flash limits. Other attempts consume the budget but do not enter the figure. All six searches reached 200 feasible trials. Searches required 202--230 attempts and 12.6--19.5 recorded trial hours. Appendix~\ref{app:cs3-proposals} reports measurement passes, timing boundaries, API usage and proposal handling.

\begin{figure}[t]
\centering
\includegraphics[width=\columnwidth]{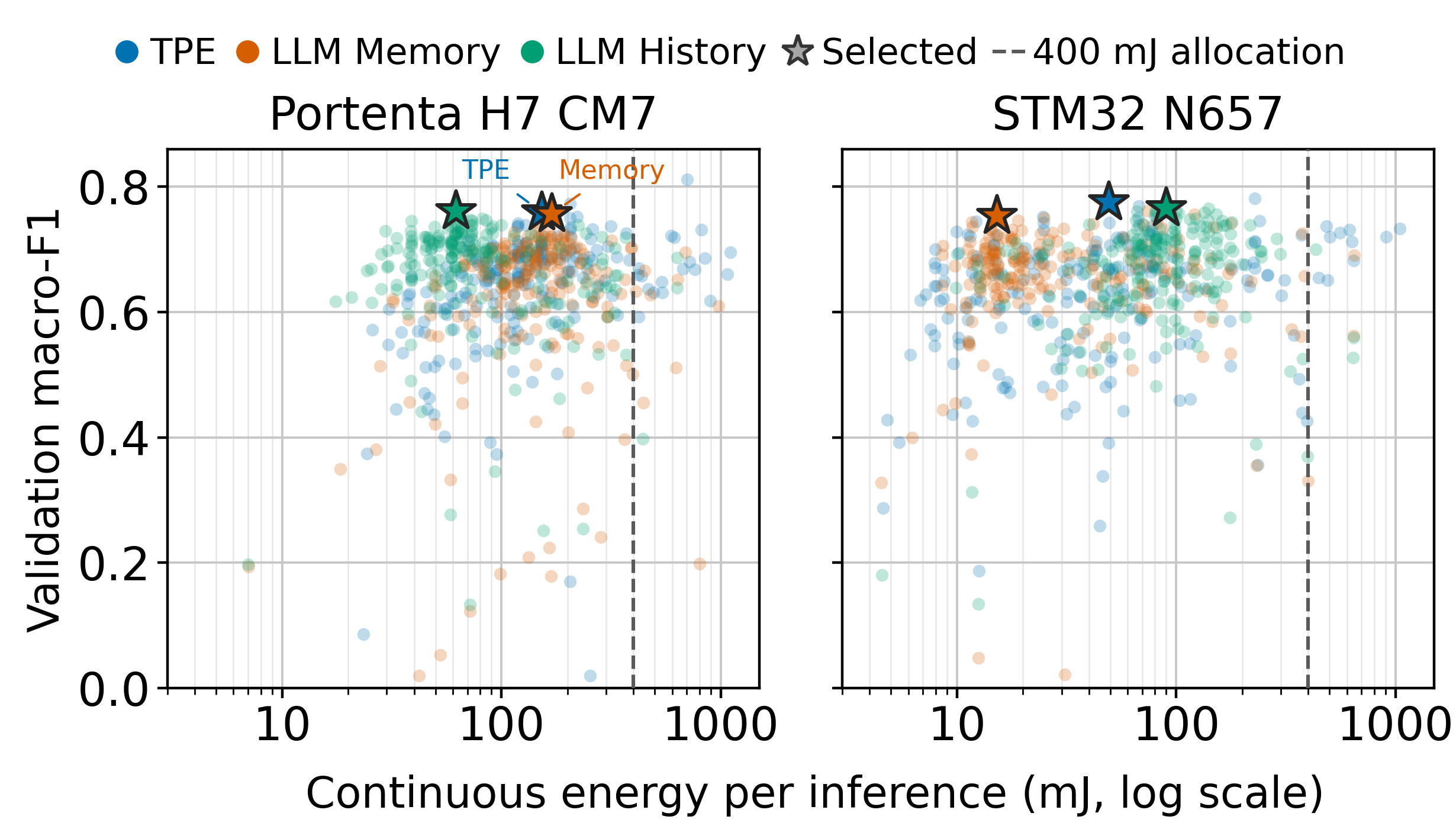}
\vspace{-1.0em}
\caption{Audio search outcomes with 200 feasible trials per strategy, including repeated configurations. Stars mark the highest score $S$. Energy uses a log scale. Dashed lines mark the 400~mJ scoring allocation, not a feasibility cutoff.}
\Description{Two side-by-side quality-energy scatter plots, with Portenta on the left and STM32 on the right. Colors identify the three proposal strategies, stars mark their score-selected candidates, and dashed vertical lines mark the 400 millijoule scoring allocation. TPE has the highest individual score on STM32. History has the highest on Portenta, where its selected candidate has similar validation quality to TPE's at lower continuous energy.}
\vspace{-1.0em}
\label{fig:cs3-audio-selection}
\end{figure}

\textbf{A common objective compares different candidate tradeoffs.}
Figure~\ref{fig:cs3-audio-selection} shows the candidate pools produced by the three proposers and the configurations selected by Equation~\ref{eq:cs3-audio-score}. The selected points illustrate how measured energy and validation quality jointly determine the application choice on each source board.

On Portenta, the selected History and TPE candidates achieve similar validation quality at different measured energy costs. History's selected candidate has validation macro-F1 of 0.761 at 61.7~mJ per inference. TPE's selected candidate has similar validation quality, 0.760, at 152.5~mJ. Measured energy therefore distinguishes these candidates under the application score. After separate final training, the int8 models achieve held-out fold-10 macro-F1 of 0.7275 for History and 0.6954 for TPE. Appendix~\ref{app:cs3-proposals} describes final training and reports held-out results for all six selections.

On STM32, TPE's selected candidate has validation macro-F1 of 0.776 at 49.3~mJ, while Memory's has 0.754 at 15.2~mJ. Under the configured score, the 0.022 macro-F1 difference outweighs the 0.0085 penalty for the additional 34.1~mJ, so the score selects TPE's candidate. This difference is within the run-to-run variation in Appendix~\ref{app:cs3-proposals}, so the example illustrates the score rather than a difference between proposers. Together, these examples show how the same scoring rule distinguishes similar-quality candidates by energy while allowing a quality improvement to justify a higher energy cost.

\textbf{Source energy ordering does not necessarily transfer.} These selections use measurements from their source board. Replaying all six configurations on the other board changes their energy ordering. The configuration TPE selected on Portenta ranks third of the six by energy on Portenta and last on STM32. It uses 218.7~mJ on STM32 compared with 152.5~mJ on Portenta, while the other five configurations use 3.0 to 4.9 times less energy on STM32 than on Portenta. A common board-to-board scaling factor would therefore miss the change in ordering. Replay retains each configuration's source validation quality and measures its hardware cost on the destination, allowing these selections to be checked without repeating the search. Appendix~\ref{app:measurement-context} gives worked examples of how deadline checks and target information affect model selection and replay.

\textbf{\systemname{} impact.}
The audio study connects candidate generation, application scoring and destination measurements through one evaluator. Changing the workload and proposer retains the comparison procedure, while cross-target replay shows why a source energy ranking needs to be checked on its destination. The records preserve source quality alongside destination hardware cost, so researchers can distinguish a change in the proposed candidates from a change in how those candidates execute.

\section{Discussion and Conclusion}
\label{sec:discussion-conclusion}

Selecting a configuration for periodic inference requires accounting for the energy consumed after inference finishes. The continuous-energy saving of a faster configuration can be offset by its additional waiting time, changing the preferred clock or architecture. At a fixed clock, changing the release period also changes the preferred architecture.

The three case studies examine different parts of the same model-selection decision. CS1 establishes that the schedule defines the energy objective and can change the preferred configuration. CS2 holds the workload and model family fixed while target deployment determines which candidates fit and which energy--quality tradeoffs are available. CS3 changes the workload and proposer, then uses replay to test the selected configurations on another board. A configuration ranked third by energy on Portenta ranks last on STM32, showing why a source selection needs evaluation on its destination target. Together, the studies show how the evaluator connects candidate selection to the conditions under which the candidate will run. The measurement records retain the quality used for selection alongside the target, interval and measured hardware cost. This lets researchers investigate a changed schedule, target or proposal strategy while preserving the information needed to interpret the resulting choices.

Held-out calibration shows that latency predicts continuous energy well and often preserves selections obtained with measured continuous energy. Direct scheduled fitting lowers prediction error without lowering selection regret. The evaluated predictors do not resolve every deployment choice. The shared-parameter model cannot represent the measured same-clock crossing, and CS3's cross-target ranking change rules out a common board-to-board energy scaling factor. These results motivate checking predicted choices through target measurement and replay. Future work could combine calibrated screening with selective scheduled replay, comparing selection quality under a common total evaluation budget. Extension to NPUs and other accelerators could examine how data transfer, wake-up costs and power management affect periodic model selection.

Our measurements use development boards at their USB supply, where sleep power remains high. Lower sleep power makes finishing early more favorable, so the direction of these effects depends on the deployment's power boundary. The same measurement path applies to an instrumented deployment board.

We presented \systemname{}, a modular HIL experimentation system for embedded model selection. Across two workloads and three boards, its shared evaluation and replay interfaces support studies with different targets, schedules and candidate proposers. Using this system, we showed on STM32 N657 that accounting for waiting energy can change the preferred clock and architecture, and that the release period can change which model to deploy. Calibrated estimates capture inference cost well, while scheduled replay exposes the differences that decide close choices. Together, the system and scheduling analysis support selecting configurations through measurements of their intended deployment conditions.

\begin{acks}
This research was funded by DEVCOM ARL under award \\\#W911NF1720196, NIH under award \#1P41EB028242, and NSF under award CNS \#2325956.
\end{acks}

\bibliographystyle{ACM-Reference-Format}
\bibliography{sensys_paper2027}

\clearpage
\appendix
\section{Waiting-power model fitting and transfer}
\label{app:cs1-waiting-model}

The calibration probe uses the int8 TCN architecture of trial 203 from the CS1 500~ms continuous-energy search on STM32 N657. At each of 200, 300, 400, 600 and 800~MHz, the probe measures ten continuous invocations and periodic windows of one or ten releases. It repeats these measurements over five rounds at 100 and 500~ms. All 150 measurements pass the timing, deadline and STOP-entry checks.

All periodic windows include the final waiting interval. Search and calibration runs keep the MCU awake in that interval. Direct replay also tests STOP after every inference. We call these policies \emph{awake-final} and \emph{STOP-final}. The earlier-firmware 200~ms periodic search used per-slot UART logging and is included only in the descriptive rank figure, not the model and selection analyses.

Let $W_{1,T}$ and $W_{10,T}$ be the measured energies of one- and ten-release windows, $E_{\mathrm{cont}}$ the continuous energy per inference, and $a_T=T-C_T$ the available waiting time. At each clock, we fit
\begin{align}
W_{1,T}-E_{\mathrm{cont}} &= a_TP_{\mathrm{busy}}+E_0, \\
\frac{W_{10,T}-10E_{\mathrm{cont}}-(W_{1,T}-E_{\mathrm{cont}})}{9}
 &= a_TP_{\mathrm{stop}}+K.
\end{align}
The two periods determine the slopes and intercepts. We pair corresponding rounds across the two period sessions and report the mean and standard deviation of the five fitted values. The window intercept $E_0$ is approximately 16~mJ, and $K$ ranges from $-1.35$ to $-1.00$~mJ. These are fitted residual corrections relative to continuous energy, not isolated physical wake or sleep-transition energies. They account for window effects, transitions and differences between continuous inference and inference after wake.

For ten releases, the waiting estimate combines nine STOP intervals and one awake interval
\begin{align}
\widehat P_{\mathrm{wait}} &= 0.9P_{\mathrm{stop}}+0.1P_{\mathrm{busy}}, \\
\widehat b &= 0.9K+E_0/10.
\end{align}
The waiting parameters are fitted to one probe architecture and shared by clock to test their transfer across architectures through Equation~\ref{eq:crest-window-energy-current}. The held-out continuous-energy fits use multiple calibration architectures, while the direct scheduled fits use scheduled-energy labels. To estimate the effect of sleeping in the final interval, we apply
\begin{equation}
E_{\mathrm{stop,est}}=E_{\mathrm{rel}}
-0.1\bigl[(T-C)(P_{\mathrm{busy}}-P_{\mathrm{stop}})-K\bigr],
\end{equation}
where $E_{\mathrm{rel}}$ is the measured awake-final energy per release.

\subsection{Reference coefficients for the timing check}

The timing-substitution checks use the latency predictor calibrated from the probe measurements
\begin{equation}
\begin{split}
\widehat E_{\mathrm{lat}}(x,T)={}&C(x)\widehat P_{\mathrm{cont}}(f_x)\\
 &+(T-C(x))\widehat P_{\mathrm{wait}}(f_x)+\widehat b(f_x).
\end{split}
\label{eq:crest-ranking}
\end{equation}
At each clock, we average energy divided by harness duration over ten continuous probe observations. Candidate duration uses the same harness definition. This predictor and the measured-continuous predictor use the same waiting parameters and correction, so only the continuous term differs between them. Both retain the probe's DWT-based waiting fit.

\begin{table}[t]
\centering
\caption{Constants for the calibrated latency predictor. Powers are in mW and corrections are in mJ per release.}
\small
\begin{tabular*}{\columnwidth}{@{\extracolsep{\fill}} r r r r @{}}
\toprule
Clock (MHz) & $\widehat P_{\mathrm{cont}}$ & $\widehat P_{\mathrm{wait}}$ & $\widehat b$ \\
\midrule
200 & 920.70 & 791.89 & 0.3708 \\
300 & 957.53 & 792.10 & 0.4593 \\
400 & 997.03 & 801.74 & 0.4428 \\
600 & 1074.17 & 802.97 & 0.5655 \\
800 & 1190.60 & 809.51 & 0.7818 \\
\bottomrule
\end{tabular*}
\end{table}

The coefficients match the ten-release search protocol. For example, they predict 85.39~mJ per release for a 600~MHz candidate taking 16.704~ms at a 100~ms period. The measured value is 84.69~mJ. The held-out estimates in CS1 instead fit continuous energy using multiple calibration architectures, separately by clock and precision.

\subsection{Direct clock and architecture comparisons}

A separate session interleaves trial 203 at 300 and 800~MHz and trial 137 at 300~MHz, both from the 500~ms continuous-energy search. Five paired rounds cover continuous execution and both final-interval policies at 100 and 500~ms. All 75 measurements pass acceptance checks. The scheduled windows have no deadline misses and record nine or ten STOP entries as specified.

With STOP after every inference, 300~MHz saves 0.26 and 1.58~mJ per release over 800~MHz for trial 203 at 100 and 500~ms, with paired 95\% intervals of 0.13--0.38 and 0.77--2.40~mJ. The awake-final savings are 0.69 and 3.35~mJ. Trial 137 at 300~MHz changes from using 1.87~mJ more than trial 203 at 800~MHz at 100~ms to 3.25~mJ less at 500~ms. Both meet the RMSE ceiling of 1.0, with source errors of 0.912 and 0.660. Table~\ref{tab:metadata-interval-example} gives absolute energies and the same-clock architecture comparison.

The direct clock and architecture comparisons use paired measurements within one session and do not rely on transferring the probe calibration. Across the calibration and direct-comparison sessions, the fitted awake-final clock-contrast slope changed from 17.5 to 6.6~mW despite matching weights, inputs and runtime settings. Board current was higher in the later session, particularly at 300~MHz, while durations remained nearly unchanged. We therefore distinguish within-session comparisons from cross-session calibration stability.

Within the direct-comparison session, the final-interval correction predicts the policy change within 0.06--0.24~mJ per release. This check evaluates the policy correction, not the stability of the full-energy calibration across sessions.

\subsection{Effective waiting slopes in the direct session}

For each configuration and paired round in the direct session, we fit $E_{\mathrm{rel}}-P_{\mathrm{cont}}C=(T-C)P_{\mathrm{wait}}+b$ to the 100 and 500~ms STOP-policy measurements using their measured active durations. The resulting effective waiting-power slopes are $803.78\pm0.28$~mW for trial 203 at 800~MHz, $800.48\pm1.79$~mW for trial 203 at 300~MHz and $791.04\pm0.28$~mW for trial 137 at 300~MHz, reported as mean and standard deviation over five rounds. These effective slopes describe the change in release energy between the two measured periods. The same-clock difference is 9.43~mW with a paired 95\% interval of 7.47--11.39~mW. The two configurations reverse energy ordering between the measured periods of 100 and 500~ms.

To describe the measured clock difference, we fit an accounting decomposition to the direct-session measurements. For trial 203 at 100~ms, we decompose the 800-minus-300~MHz energy difference using the 300~MHz slope as the shared waiting-power reference. The continuous-energy difference is $-9.043$~mJ. Additional waiting at the shared power contributes $8.799$~mJ, the slope difference contributes $0.309$~mJ and the fitted correction difference contributes $0.193$~mJ. Their sum is the measured $0.258$~mJ difference. At 500~ms, the slope-difference contribution grows to $1.632$~mJ and the total becomes $1.582$~mJ. This decomposition uses parameters fitted to the same direct measurements.

\subsection{Repeated measurements of search-pool examples}
\label{app:cs1-repeat-sensitivity}

Twenty paired rounds at 500~ms test two search-time reversal examples using fixed artifacts. Table~\ref{tab:cs1-repeat-sensitivity} reports the paired differences. Positive release-energy differences reverse the continuous ordering. These are separate from the direct comparisons above.

\begin{table}[t]
\centering
\caption{Search-pool pair sensitivity at 500~ms, in mJ. Entries are first minus second configuration. Brackets give paired $t$ 95\% intervals over twenty rounds. Estimated STOP-final values are conditional on the fitted correction. Their intervals do not propagate parameter uncertainty.}
\label{tab:cs1-repeat-sensitivity}
\small
\setlength{\tabcolsep}{3pt}
\begin{tabular*}{\columnwidth}{@{\extracolsep{\fill}} l r r @{}}
\toprule
Difference & 203--114 & 228--131 \\
\midrule
Clocks (MHz) & 800 / 200 & 600 / 600 \\
Continuous & $-91.7$ & $-3.11$ \\
\midrule
\multicolumn{3}{@{}l}{\textit{Measured awake-final differences}} \\
Release energy & $2.49$ & $0.28$ \\
95\% interval & $[2.08,2.90]$ & $[0.14,0.43]$ \\
\midrule
\multicolumn{3}{@{}l}{\textit{Model-derived STOP-final estimates}} \\
Release energy & $-0.29$ & $0.25$ \\
Conditional interval & $[-0.70,0.12]$ & $[0.11,0.40]$ \\
\bottomrule
\end{tabular*}
\end{table}

Repeated measurements qualify the magnitude and policy dependence of these two search-time examples, which are distinct from the directly measured STOP-final comparisons above. For the 203--114 pair, the measured release-energy gap retains its direction but decreases from 11.38~mJ at search time to 2.49~mJ in the repeat session. Its estimated STOP-final interval includes zero. The 600~MHz pair 228--131 retains a resolved within-session difference of 0.28~mJ per release. These repeats support the awake-final ordering. The STOP-final values here are estimates.

\subsection{Prediction with held-out architectures}
\label{app:cs1-heldout}

The 756 observations contain 470 inference-graph identities. Identity includes filters, kernel size, the ordered dilation list, skip connections, normalization and input shape. Dropout and training batch size are excluded. Copies across pools, frequencies and precision modes remain in the same fold. Each of ten deterministic repetitions assigns graphs to five folds. Fits use the other four folds and give each calibration graph equal total weight. Proportional and affine continuous fits are separate for each pool, clock and precision. Both retain the original waiting calibration when predicting release energy. Direct scheduled affine fits use release-energy labels and continuous duration from the same calibration subset.

A fitted group requires at least three calibration graphs and two distinct durations. All estimates are compared on the same supported observations, giving 7,502 held-out predictions over 753 distinct observations. Three sparse-group observations have no supported split. Across the 1,357 supported group fits, the calibration set contains 3 to 47 distinct architectures, with a median of 20. This evaluation tests transfer to held-out architectures at the available calibration scale. Direct scheduled fits use unconstrained slopes. The median group continuous RMS errors are 0.61~mJ for affine and 1.13~mJ for proportional fitting, or 1.69\% and 2.34\% of group mean energy. Table~\ref{tab:cs1-heldout-prediction} instead aggregates squared errors over all held-out predictions.

Among the 700 int8 observations, release-energy RMS errors are 4.10~mJ for fitted proportional, 4.04~mJ for fitted affine, 3.59~mJ using measured continuous energy and 3.18~mJ for direct scheduled fitting.

A separate transfer check excludes an entire pool and every graph appearing in it from calibration. Frequency- and precision-specific continuous fits then predict the excluded pool using the frozen waiting model. This covers 700 int8 observations. Affine continuous RMS error is 1.80~mJ and release-energy RMS error is 4.30~mJ, compared with 3.59~mJ when continuous energy is measured. The float observations have no calibration counterpart in another pool. This comparison changes the acquisition pool as well as the architectures. Direct scheduled fits are evaluated within pools, where period and policy are fixed.

\subsection{Matched-latency replay and transfer}
\label{app:cs1-transfer-replay}

The 400~MHz pairs below come from the 500~ms periodic search and are distinct from the continuous-search pair 228--131 above. Trial numbers are local to each search, so the two references to trial 228 identify separate trials. The initial pair, periodic trials 110 and 123, was first measured in five rounds and then repeated with the same saved binaries in twenty rounds. Six additional pairs were randomly selected within frequency strata with seed 20261102, using search-time latency differences below 2\% and excluding duplicate inference architectures. Each selected pair has two different architectures, and no architecture is shared between these six pairs. A targeted pair at 600~MHz compares operator composition at similar latency, using a wider skip-connected model and a narrower model without skip connections. It is separate from the random panel.

Each round randomizes pair-block order and measurement order within each block. A reference configuration is measured once per round of the random panel. All 380 measurements pass acceptance checks, with ten releases, nine STOP entries and no deadline misses in each scheduled window. The replayed 600~MHz random pair has a measured latency separation of about 2.2\%. Table~\ref{tab:cs1-transfer-pairs} reports all eight pairs, including those without resolved release-energy differences. Comparisons are paired within rounds.

\begin{table}[t]
\centering
\caption{Replay differences, first candidate minus second. Duration is continuous harness time in ms. Energies and paired 95\% intervals are mJ. The initial pair has twenty rounds and the others ten. C and S identify continuous and periodic source pools.}
\label{tab:cs1-transfer-pairs}
\scriptsize
\setlength{\tabcolsep}{2pt}
\begin{tabular*}{\columnwidth}{@{\extracolsep{\fill}} l r r r l @{}}
\toprule
Pair & MHz & $\Delta C$ & $\Delta E_{\mathrm{cont}}$ & $\Delta E_{\mathrm{rel}}$ [95\% interval] \\
\midrule
S 110 / 123 & 400 & 0.239 & 0.010 & $-6.778$ [$-7.024,-6.531$] \\
C 163 / 195 & 200 & $-0.327$ & $-0.478$ & $-0.212$ [$-0.566,0.143$] \\
S 74 / 167 & 300 & $-1.002$ & $-2.048$ & $-1.571$ [$-1.846,-1.296$] \\
S 228 / 243 & 400 & 0.308 & 0.473 & $-3.623$ [$-3.816,-3.430$] \\
C 82 / 134 & 600 & 7.001 & 6.273 & 1.155 [0.878,1.433] \\
C 83 / 135 & 800 & 0.186 & 0.328 & 0.108 [$-0.100,0.316$] \\
S 171 / 183 & 400 & $-0.314$ & $-0.174$ & 0.248 [$-0.032,0.527$] \\
Targeted & 600 & 0.379 & 0.863 & 0.943 [0.656,1.229] \\
\bottomrule
\end{tabular*}
\end{table}

The random-panel release differences at 300~MHz, at 400~MHz for 228/243 and at 600~MHz remain resolved after Holm adjustment across the six comparisons. The displayed intervals are unadjusted paired intervals. For 110/123, the measured-continuous predictor leaves a mean residual of $-6.60$~mJ with interval $[-6.84,-6.36]$~mJ. For 228/243 it leaves $-3.85$~mJ with interval $[-4.06,-3.63]$~mJ. These residual intervals condition on the frozen calibration. The comparisons test prediction transfer rather than assigning the residual to a particular execution phase.

We also compare active timing using DWT cycles divided by nominal clock in both execution modes. Under scheduled execution, 110 is 0.69~ms slower than 123, while 228 is 0.96~ms faster than 243. Substituting these scheduled differences into the frozen latency model predicts release differences of $+0.13$ and $-0.19$~mJ, respectively. This timing substitution does not recover the observed gaps under those coefficients.

\subsection{Paired replay under both final-interval policies}
\label{app:cs1-final-policy-replay}

Ten further rounds compare periodic-search trials 110, 123, 228 and 243 at 400~MHz under continuous execution and both final-interval policies. Each round includes a reference configuration and randomizes pair-block and measurement order. All 150 measurements pass acceptance checks. The scheduled windows contain ten releases at 500~ms, record the expected STOP entries and have no deadline misses.

Continuous and awake-policy runs reuse the saved images. STOP variants are compiled from copies of the saved projects using the existing sleep and wake routine in the final interval. Weight hashes match the originals. Table~\ref{tab:cs1-final-policy-replay} gives paired differences. Both model pairs retain their direction in every round under both policies.

\begin{table}[t]
\centering
\caption{Energy differences at 400~MHz and 500~ms, first trial minus second, in mJ per release. Brackets give paired 95\% intervals over ten rounds.}
\label{tab:cs1-final-policy-replay}
\small
\begin{tabular*}{\columnwidth}{@{\extracolsep{\fill}} l l r @{} }
\toprule
Pair & Final interval & Difference [95\% interval] \\
\midrule
110 / 123 & Awake & $-6.46$ [$-6.55,-6.37$] \\
110 / 123 & STOP & $-0.21$ [$-0.28,-0.14$] \\
228 / 243 & Awake & $-3.73$ [$-3.96,-3.49$] \\
228 / 243 & STOP & $-5.81$ [$-5.92,-5.71$] \\
\bottomrule
\end{tabular*}
\end{table}

These comparisons measure the complete execution of each compiled policy variant, rather than isolating the energy of the final interval. Comparing the STOP-final and awake-final builds changes the 110/123 contrast by $+6.25$~mJ, with interval $[6.17,6.34]$~mJ, and the 228/243 contrast by $-2.09$~mJ, with interval $[-2.37,-1.81]$~mJ. At the individual-model level, the STOP variant increases 110's energy by 3.17~mJ and decreases 123's by 3.08~mJ.

For 228/243, continuous energy is 0.57~mJ higher for 228 in this experiment. DWT cycles divided by nominal clock give a continuous active-time difference of $+0.37$~ms, compared with $-0.91$~ms under both scheduled policies. The scheduled timing difference therefore has the energy difference's direction, while remaining nearly unchanged as the energy gap grows from 3.73 to 5.81~mJ. Substituting that timing into the previous shared latency coefficients gives about $-0.18$~mJ under either policy. Both timing checks use the continuous calibration coefficients without refitting to scheduled measurements.

\subsection{Waiting calibration and prediction residuals}

Table~\ref{tab:cs1-waiting-powers} reports the fitted waiting powers used by the continuous-derived estimates. They include the ST-LINK debugger and other components powered through the board's USB supply, which continue to contribute to board power when the MCU enters STOP.

\begin{table}[t]
\centering
\caption{Fitted waiting powers for the STM32 N657 probe architecture. Values are means and standard deviations over five paired rounds. Powers cover the whole board.}
\label{tab:cs1-waiting-powers}
\small
\begin{tabular*}{\columnwidth}{@{\extracolsep{\fill}} r r r @{}}
\toprule
Clock & Busy wait & STOP \\
(MHz) & (mW) & (mW) \\
\midrule
200 & $849\pm5$ & $785.5\pm1.3$ \\
300 & $852\pm3$ & $785.5\pm1.4$ \\
400 & $860\pm4$ & $795.3\pm2.5$ \\
600 & $871\pm6$ & $795.4\pm4.0$ \\
800 & $906\pm8$ & $798.8\pm2.6$ \\
\bottomrule
\end{tabular*}
\end{table}

\begin{figure}[t]
\centering
\includegraphics[width=0.8\columnwidth]{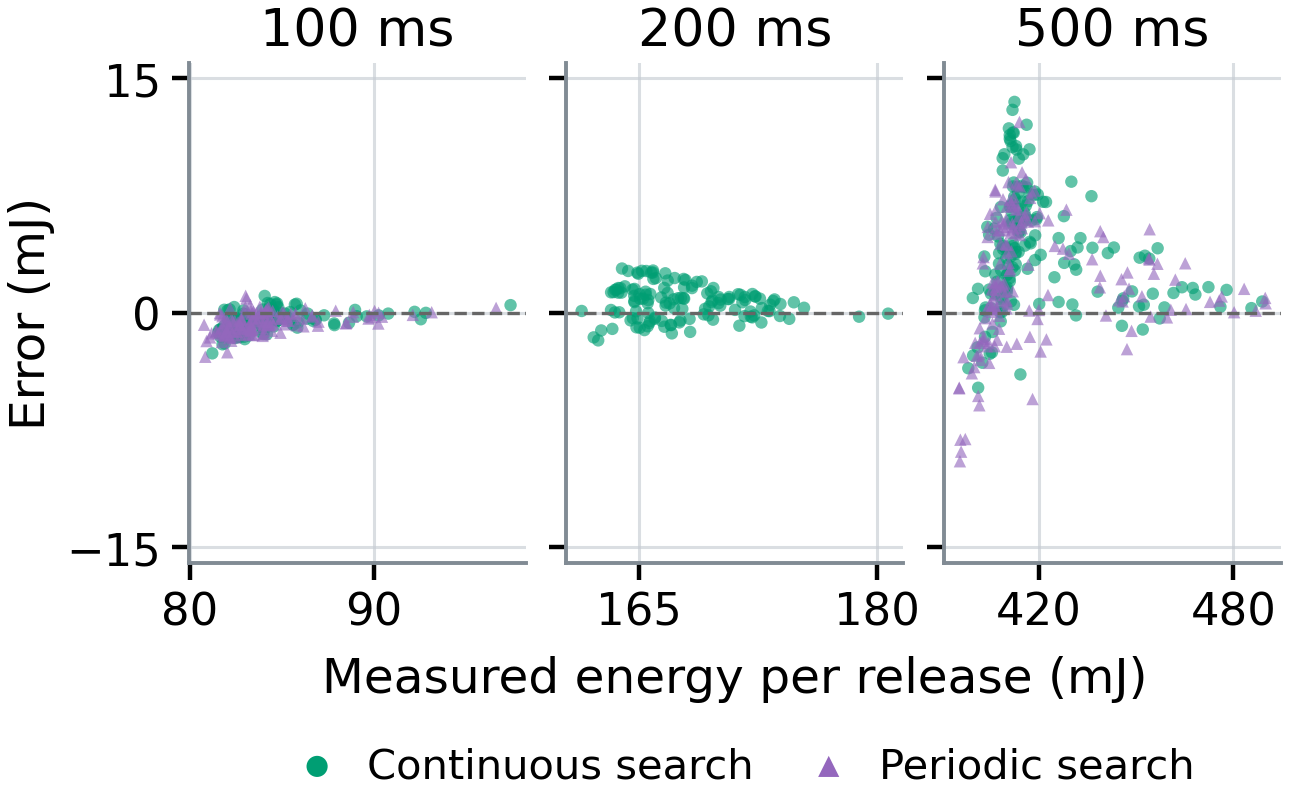}
\caption{Measured minus predicted energy per release for the measured-continuous predictor across five search pools.}
\Description{Prediction residuals at 100, 200 and 500 ms. Green circles represent continuous-search candidates and purple triangles represent periodic-search candidates.}
\label{fig:cs1-model-transfer}
\end{figure}

Figure~\ref{fig:cs1-model-transfer} shows measured minus predicted energy per release for the measured-continuous predictor across the five eligible search pools. RMS error is approximately 0.9~mJ at 100~ms, 1.2~mJ at 200~ms and 4.5--5.5~mJ at 500~ms.

\section{CS1 model selection under periodic execution}
\label{app:cs1-selection}

We compare selection objectives within each eligible search pool under common validation RMSE ceilings of 0.65, 0.75 and 1.0. A selected candidate's observed regret is its measured energy per release minus the lowest recorded energy per release among eligible candidates at that ceiling. This reference is an observed minimum, not an independently established minimum of expected energy. Each observation contributes one energy value, so measurement noise can make the reference optimistic. All estimates use the same eligible observations and reference within each comparison. The comparison uses trained validation quality and search-pool hardware measurements under the awake-final search protocol. Appendix~\ref{app:cs1-waiting-model} describes the protocol and analysis coverage.

\subsection{Selection with held-out calibration}

For the architecture-separated evaluation in Appendix~\ref{app:cs1-heldout}, selection is performed within each held-out fold and pool under the same RMSE ceilings of 0.65, 0.75 and 1.0. Each comparison uses the same eligible observations for every estimate and requires at least two candidates. Across repeated splits this gives 622 comparisons. The reference is the lowest measured release energy among the eligible observations in that subset.

Mean observed regret is 1.72~mJ for the continuous affine estimate, 1.71~mJ for proportional fitting, 1.64~mJ for measured continuous energy and 1.73~mJ for direct scheduled fitting. The continuous affine and measured-continuous estimates select equal measured energy in 542 comparisons. Measured continuous energy selects lower energy in 58, while the affine estimate selects lower energy in 22. Selecting by continuous energy alone gives mean observed regret of 2.64~mJ and selecting by latency alone gives 2.77~mJ. Compared with continuous energy alone, the measured-continuous estimate selects lower energy in 258 comparisons, equal energy in 255 and higher energy in 109. The comparisons reuse observations across folds and quality ceilings, so these counts summarize selection behavior rather than independent search runs.

In the excluded-pool transfer check, the five pools and three quality ceilings give fifteen comparisons. Mean observed regret is 2.93~mJ for the affine estimate, 2.61~mJ using measured continuous energy and 4.13~mJ for continuous-energy selection. Selected energy is equal in twelve cases and lower with measured continuous energy in three, by at most 2.10~mJ. These results use the same measured-energy reference as the preceding regret analysis.

\section{Measurement context in deployment decisions}
\label{app:measurement-context}

This appendix illustrates feasibility and provenance checks and documents the whole-board supply boundary. Table~\ref{tab:metadata-interval-example} in \S\ref{subsec:cs1-schedule} reports the direct experiment's absolute energy values.

\subsection{Feasibility and target provenance}

Trial 115 from the 100~ms continuous search produces a valid energy measurement of 91.27~mJ per release, but inference takes 104.81~ms and all ten releases miss their deadlines. The record retains the measurement and failed timing checks while excluding this candidate from selection for the 100~ms schedule. Cross-target replay likewise keeps source deadline results separate from destination feasibility. For the STM32 History selection, trial 187, Portenta energy is measured but its unsupported deadline-miss count remains unavailable rather than inheriting the STM32 pass.

\subsection{Board energy in sleep and active states}

To examine board energy in different execution states, we ran a fixed-duration phase-energy probe on the evaluated target configurations. Each trial measured one second while the DUT entered its sleep path or executed an integer or floating-point arithmetic loop. Table~\ref{tab:phase-energy-probe} reports the mean over 50 trials.

The measurements use the development-board supply boundary described in \S\ref{subsec:impl-hil-harness}. They include all components powered through that input and characterize whole-board energy rather than isolated MCU power. All measurements were taken indoors, and we did not vary ambient temperature. The integer and floating-point loops run for the same duration and complete different amounts of work, so their energies are not comparable per operation. This fixed-duration characterization is separate from the clock-specific waiting-power calibration in Appendix~\ref{app:cs1-waiting-model}. Its sleep-path energies are not used as fitted STOP powers.

\begin{table}[t]
\centering
\caption{One-second phase-energy probe. Values are mean board energy in mJ over 50 trials and include all components powered through the instrumented supply.}
\label{tab:phase-energy-probe}
\footnotesize
\setlength{\tabcolsep}{4pt}
\renewcommand{\arraystretch}{1.08}
\begin{tabular*}{\columnwidth}{@{\extracolsep{\fill}} l r r r @{}}
\toprule
\textbf{Target} &
\textbf{Sleep [mJ]} &
\textbf{Active int [mJ]} &
\textbf{Active FP [mJ]} \\
\midrule
Nano 33 BLE & 24.8 & 46.7 & 43.9 \\
Portenta M4 & 940.2 & 993.4 & 976.7 \\
Portenta M7 & 932.6 & 1045.5 & 1020.3 \\
STM32 N657 & 823.1 & 892.5 & 905.1 \\
\bottomrule
\end{tabular*}
\end{table}

\section{Training-budget sensitivity}
\label{app:training-budget}

We examined training-budget sensitivity using 50 configurations selected from both the search-time Pareto front and dominated configurations of the CS2 Portenta H7 CM7 measured-energy search, including 22 float and 28 int8 configurations. Each configuration was trained with seed 17 for 300 epochs using the same OxIOD training and validation splits, batch size, and training procedure. We evaluated the checkpoint with the lowest validation loss available within budgets of 15, 30, 55, 100, 200, and 300 epochs. Quality was measured using host TFLite validation error, while search-time inference energy remained fixed. We also reconstructed the original patience-40 early-stopping policy along each trajectory. Both checkpoint policies produced identical results at the 55-epoch budget.

\begin{figure}[!htbp]
    \centering
    \includegraphics[width=0.78\columnwidth]{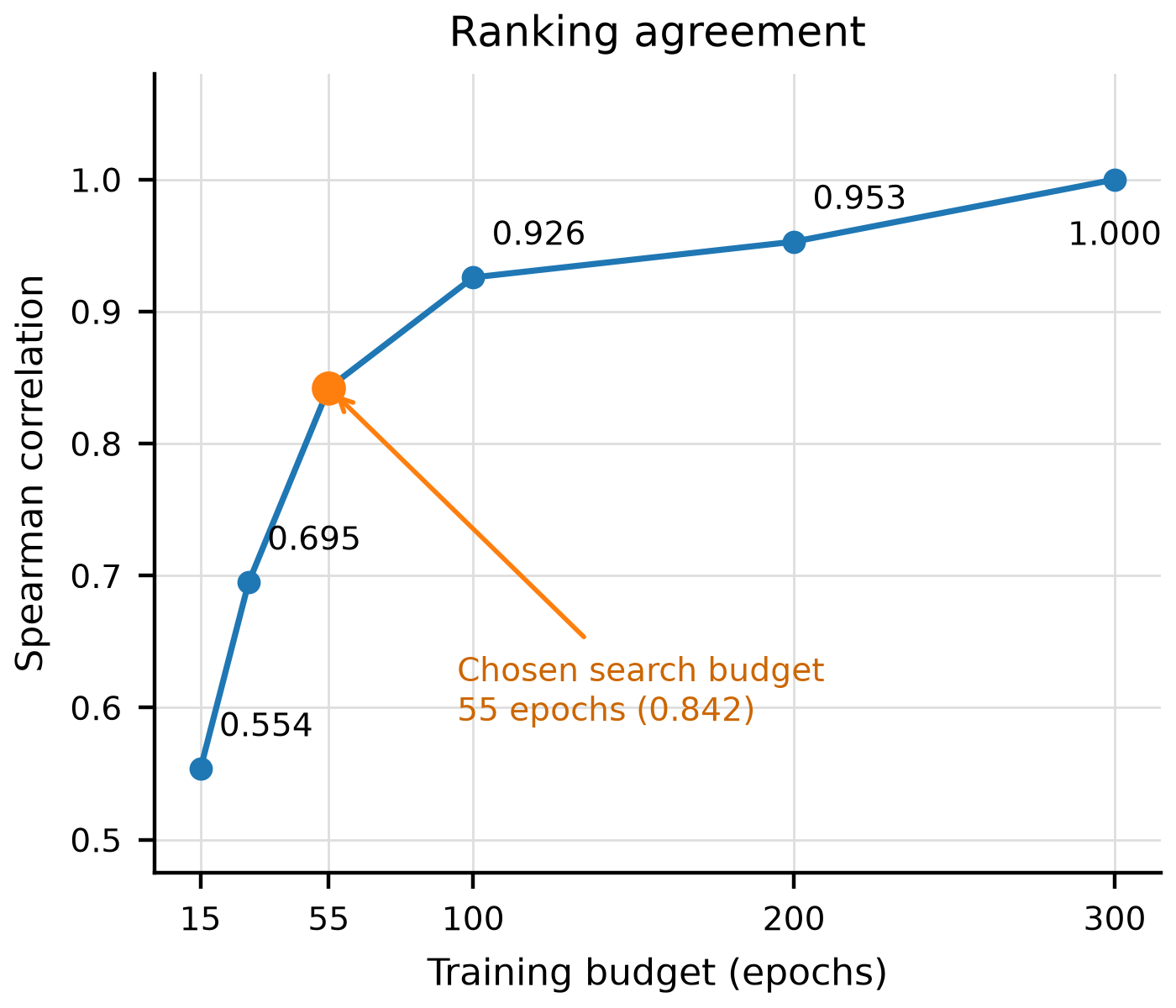}\\[0.5em]
    \includegraphics[width=0.78\columnwidth]{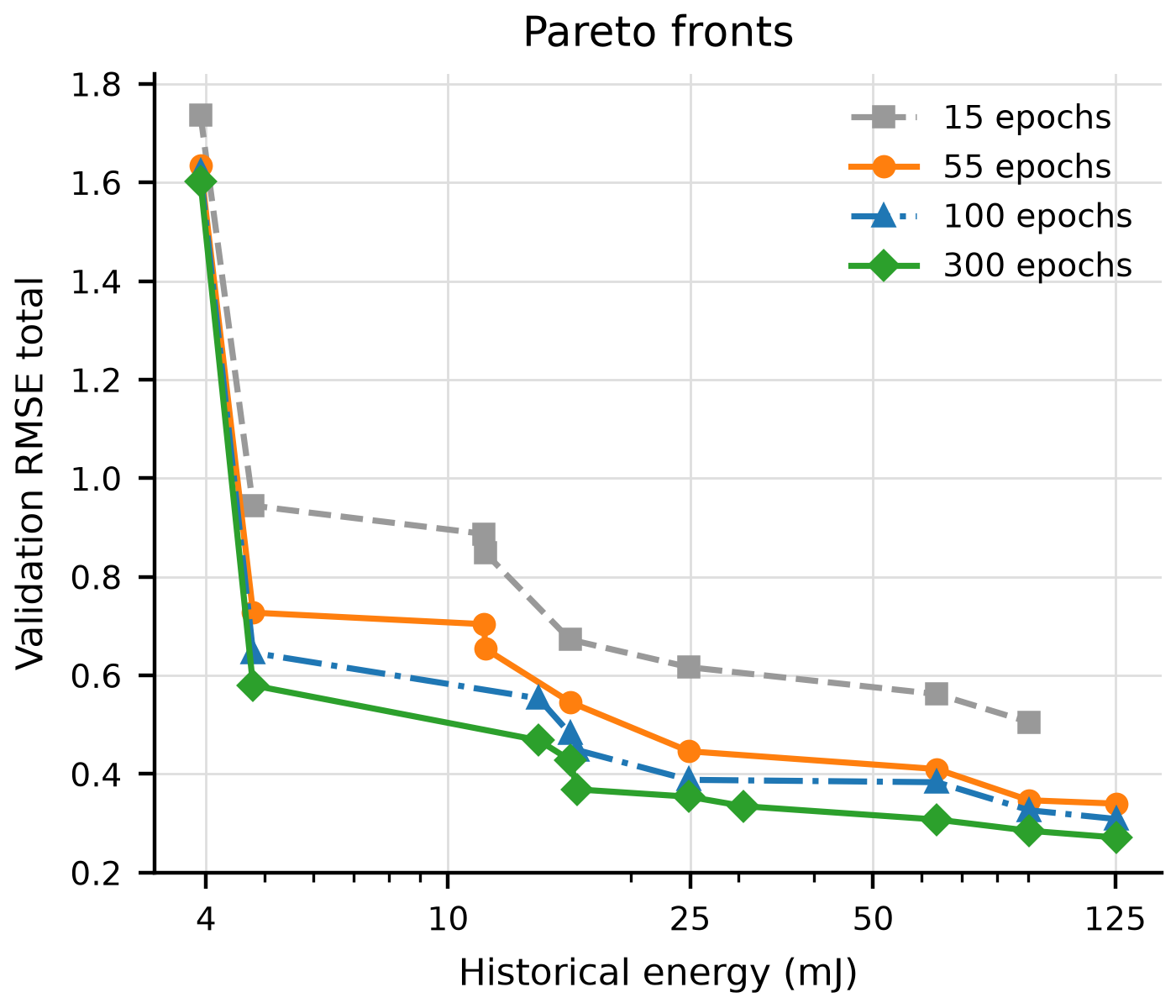}
    \caption{Training-budget sensitivity for 50 CS2 configurations. The top panel shows validation-error rank correlation with the 300-epoch reference, with 55 epochs highlighted. The bottom panel shows Pareto fronts using the best validation-loss checkpoint within each budget. Both panels use seed 17 and fixed search-time Portenta CM7 energy. Lines guide the eye.}
    \Description{Two vertically stacked plots show ranking agreement increasing with training budget and the corresponding validation-error versus energy Pareto fronts.}
    \label{fig:cs2-training-budget}
\end{figure}

Figure~\ref{fig:cs2-training-budget} shows how training budget affects agreement with the 300-epoch reference and the resulting selection tradeoffs. Spearman rank correlation increased from 0.554 at 15 epochs to 0.842 at 55 epochs, 0.926 at 100 epochs, and 0.953 at 200 epochs under continued training. Across the 50 configurations, cumulative training time, including checkpoint saving and logging, was approximately 4.8 hours through epoch 55 and 24.4 hours through epoch 300. These timings represent prefixes of the same trajectories and exclude subsequent checkpoint evaluation and hardware measurement. We use the 55-epoch budget to limit per-candidate training cost. The longer budgets characterize the remaining selection sensitivity on this panel.

\FloatBarrier
\section{Candidate evaluation cost}
\label{app:time-study}

We profiled fixed OxIOD configurations to separate training and host-side evaluation from the additional cost of preparing, deploying and measuring candidates on hardware. This study measures evaluation cost, not adaptive-search or candidate-proposal overhead.

\begin{figure*}[!t]
    \centering
    \includegraphics[width=0.85\textwidth]{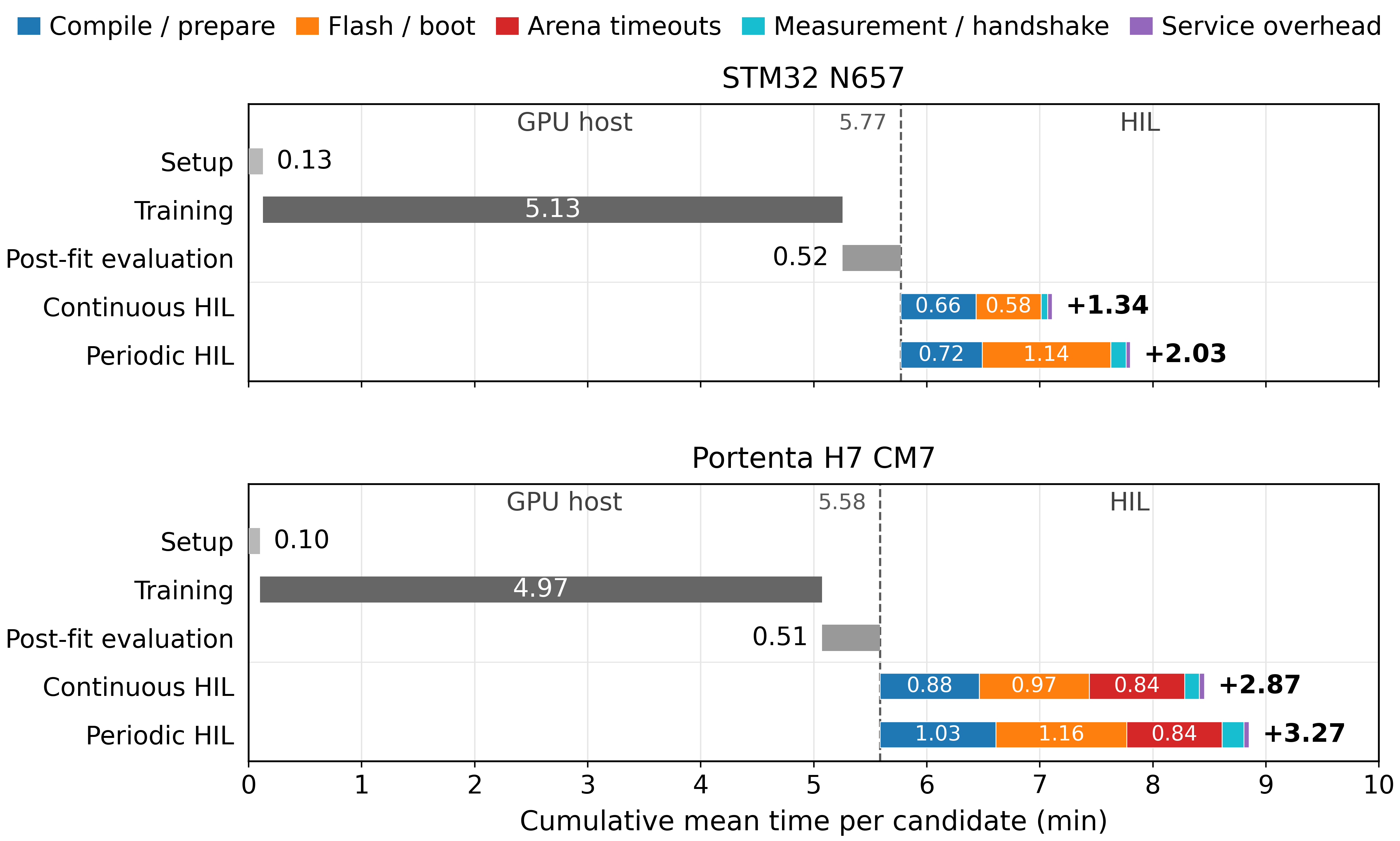}
    \caption{Mean time per candidate for the fixed OxIOD panels. Gray bars show shared GPU-host work. Colored bars show additional continuous or 200~ms periodic HIL costs, with totals in bold. Values are minutes. Bars group costs by component, not execution order. Timeout segments include failed arena probes' measurement and protocol handling. Their compilation and programming remain in the preparation and flashing segments.}
    \Description{Two timing panels for STM32 and Portenta M7. Shared host work totals 5.77 and 5.58 minutes, respectively. Continuous and periodic HIL add 1.34 and 2.03 minutes on STM32, and 2.87 and 3.27 minutes on Portenta. Portenta includes 0.84 minutes of arena-probe timeout handling in both modes.}
    \label{fig:appendix-time-study}
\end{figure*}

\subsection{Setup and timing boundaries}

For each of STM32 N657 and Portenta H7 CM7, we selected ten unique configurations at evenly spaced positions along the CS2 measured-energy Pareto front, including both endpoints. The STM32 panel contained eight float and two int8 configurations. The Portenta panel contained ten float configurations. These were separate target-specific panels. Within each panel, architecture, precision and applicable clock settings were held fixed across continuous HIL, 200~ms periodic HIL, FLOPs-proxy and memory-traffic-proxy evaluation.

All 80 evaluations used the OxIOD splits, batch size 256, seed 42 and a 55-epoch training budget, and completed without candidate-level failures or rejections. Proxy evaluation performed training and host-side quality evaluation without firmware compilation or hardware requests. Continuous HIL measured ten invocations. Periodic HIL collected both a ten-invocation continuous measurement and a ten-release periodic measurement. Both proxy quantities were computed in all conditions.

The optimizer client ran on an NVIDIA RTX A5000 GPU host, with the HIL server on a separate workstation connected to the DUT and harness. Each condition evaluated its ten candidates sequentially within one client process and, where needed, one server process. Nested elapsed-time spans separate loading and setup, training with validation and checkpoint saving, post-fit evaluation, and the complete HIL request without double counting. HIL includes export, backend preparation, compilation, programming, boot, measurement, protocol handling and communication. Startup and lazy dataset initialization are included and amortized across each panel.

Figure~\ref{fig:appendix-time-study} pools the shared host components over 40 evaluations per target and averages each HIL mode over ten. The bars group costs by component rather than execution order. Hardware checks occur before training. Each total combines pooled host means with the corresponding HIL mean. Internal arena probes and their timeouts remain included in the reported costs.

\FloatBarrier
\subsection{Results and scope}

Shared host work averaged 5.77~min per candidate on STM32 and 5.58~min on Portenta. Training, including validation and checkpoint saving, accounted for approximately 89\% of this work, averaging 5.13 and 4.97~min, respectively. Post-fit evaluation added 0.52 and 0.51~min. Loading and setup added 0.13 and 0.10~min. Proxy calculation was included in setup and was small relative to training.

Continuous and periodic HIL added 1.34 and 2.03~min per candidate on STM32, compared with 2.87 and 3.27~min on Portenta. STM32 spent approximately 35~s in export and ST Edge AI preparation, followed by 4.8~s of firmware build time for continuous evaluation or 8.0~s for periodic evaluation. Portenta's tensor-arena search rebuilt, uploaded and tested successive memory sizes. Each Portenta condition used 53 base attempts, comprising 39 successful measurements and 14 undersized-arena probes that timed out. Periodic HIL added ten follow-up measurements. The unsuccessful probes' measurement and protocol handling averaged 50.5~s per candidate in both modes, with their compilation and programming accounted for separately. These internal probes explain part of the backend overhead. All candidate evaluations completed successfully. STM32 used ten continuous or twenty periodic phase attempts without an arena sweep.

Each condition was run once on its selected panel. The means characterize these candidates, hosts, training budget and backend implementations, not variation across repeated campaigns or an intrinsic speed difference between MCUs. All candidates passed the checks, so the study does not measure training time saved by early rejection. Manual board setup and candidate proposal are excluded. This OxIOD breakdown is not extrapolated to CS3's audio searches.

\FloatBarrier

\section{CS3 proposal policies and evaluation}
\label{app:cs3-proposals}

This appendix describes how the Memory and History policies introduced in \S\ref{subsec:impl-search-failure} generate candidates for CS3. These are complete proposal policies evaluated through a common deployment evaluator, not an isolated comparison of summarization or a test of general optimizer superiority. We detail the information supplied to each policy, the rules used to accept proposals, and the campaign settings. The final subsection reproduces the fixed instruction texts and describes the context added to each request.

\subsection{Shared proposal contract}

CS3 uses one categorical DS-CNN search space and one application score for all three proposal strategies. Optuna manages the study, trial states, and objective evaluation. Its TPE sampler can select parameters within a trial. An LLM proposer instead returns explicit raw parameter dictionaries. The client validates those dictionaries against the declared search space and the policy's proposal rules before queueing accepted candidates as Optuna trials. The shared evaluator then supplies training quality, feasibility, and measured hardware cost. Rejected proposals are recorded without consuming trial attempts.

Both LLM policies receive the search space, task and input description, target resource limits, training and quantization settings, score definition, trial outcomes, pending candidates, and remaining budgets. They use OpenAI's \texttt{gpt-6-luna} with medium reasoning effort. Each of the four LLM campaigns evaluates the same 15 fixed warmup architectures on its target before requesting batches of up to five candidates. The model receives a structured output contract requiring exact search-space keys. The client checks returned values locally and records proposal source, request identity, validation failures, and random fallback. For CS3, fold 10 remains reserved for final testing after candidate selection.

\subsection{Memory policy}

Memory uses a summarization scheme inspired by context compaction. It keeps recent results in the candidate prompt and summarizes older ones. Each request includes the 15 most recent terminal trials, five strong-candidate anchors, and a generated knowledge base. The summarizer is instructed to retain the conditions and uncertainty behind older observations.

The first summary is produced after 30 terminal trials and covers trials 0 to 14. The summarizer then updates every 15 trials, leaving the latest 15 for direct feedback. Up to 15 older trials awaiting summarization can also appear in the request. Python computes the numeric aggregates supplied to the summarizer. The knowledge base holds at most 12 findings and 6,000 characters.

The Memory candidate prompt asks for exact raw parameters and uses the generated findings with their recorded conditions and uncertainty. The separate summary prompt asks the model to preserve measurements and exceptions when incorporating new trials.

\subsection{History policy}

History sends a compact columnar record of every finished trial, including parameters, decoded architecture identity, state, quality, score, continuous energy, latency, resource use, and failure information. Pending candidates are included separately. Missing measurements remain unavailable rather than becoming zero. The request also supplies architecture coverage, eligible feasible parents, and an allocation between local changes and exploration. A local proposal names a parent and describes a change. An exploration proposal follows an assigned coverage target while those targets remain active.

The History policy rejects raw and decoded duplicates and can request up to two refills for unfilled batch slots while retaining already accepted candidates. It does not generate a summary. Memory and History vary the evidence shown to the model, proposal instructions, validation, and request handling. The comparison therefore evaluates complete proposal policies and does not isolate the effect of summarization.

\subsection{Campaign settings and measurements}

Random fallback supplies randomly sampled candidates when the LLM proposal path does not supply an accepted candidate. These candidates follow the same training, deployment, measurement and scoring path as accepted LLM proposals. The Memory campaigns include 50 fallback attempts on Portenta and seven on STM32. Neither History campaign used random fallback.

All six CS3 searches score continuous energy per inference. On Portenta, all three searches obtained this measurement from continuous inference. The LLM searches also collected periodic measurements in a second pass. The STM32 searches collected continuous and periodic measurements and scored continuous energy. Host-side training supplied validation quality, while HIL measured configuration deployment cost.

The additional periodic passes are included in recorded trial times, so the Portenta timing comparison includes different measurement work across strategies. Each search outcome comes from one campaign per target and policy. For the STM32 TPE campaign, 12 exact parameter configurations have repeated evaluations, totaling 28 observations. Pooling the within-configuration sums of squared validation macro-F1 deviations and dividing by the sum of their degrees of freedom gives a standard deviation of 0.0466.

\subsection{Final training and held-out evaluation}

After search, each selected configuration is rebuilt and trained separately using the training folds, with fold 9 providing validation loss for checkpoint selection and early stopping. The selected checkpoint is exported with int8 PTQ and evaluated using the host TFLite interpreter on held-out fold 10. These test results are separate from the validation quality used in search and inherited during replay. In TPE, Memory and History order, final test macro-F1 is 0.6954, 0.6446 and 0.7275 on the Portenta selections, and 0.7132, 0.6741 and 0.6886 on the STM32 selections.

\subsection{Scoring scale and sensitivity}

We use 400~mJ as a reference energy scale and 0.10 as the quality--energy tradeoff weight. An inference using 400~mJ receives a penalty of 0.10 in macro-F1 units. These values define the common scoring policy. The energy scale sets the score penalty, while latency and memory are checked separately for feasibility.

Re-scoring uses the trials each search collected (Table~\ref{tab:cs3-weight-sensitivity}), so it does not show which candidates a search under another weight would propose. Across these collected trials, the highest-scoring observations remain in the History pool on Portenta and the TPE pool on STM32 for energy weights from 0.033 to 0.248, even as the selected trials change.

\begin{table}[t]
\centering
\caption{CS3 selections at the 0.10 energy weight and the range of weights over which each search keeps the same selection. Each search re-scores its 200 feasible trials.}
\label{tab:cs3-weight-sensitivity}
\small
\setlength{\tabcolsep}{3pt}
\begin{tabular*}{\columnwidth}{@{\extracolsep{\fill}} l l r r l @{}}
\toprule
Board & Proposer & Macro-F1 & Energy (mJ) & Weight range \\
\midrule
Portenta & TPE & 0.760 & 152.5 & 0.0995--0.207 \\
Portenta & Memory & 0.757 & 169.7 & 0--0.209 \\
Portenta & History & 0.761 & 61.7 & 0--0.269 \\
STM32 & TPE & 0.776 & 49.3 & 0.013--0.339 \\
STM32 & Memory & 0.754 & 15.2 & 0.047--1 \\
STM32 & History & 0.766 & 89.7 & 0--0.1005 \\
\bottomrule
\end{tabular*}
\end{table}

\subsection{Search time and API usage}

\begin{table}[t]
\centering
\caption{CS3 recorded campaign effort and API usage. Each cell lists Portenta / STM32. Trial time sums recorded attempt durations, including failed and pruned attempts. Token counts are millions, include Memory summary requests, and cover saved responses. Portenta LLM runs also collected a periodic pass. Trial times therefore include different measurement work and do not establish comparative optimizer efficiency.}
\label{tab:cs3-search-cost}
\footnotesize
\setlength{\tabcolsep}{2pt}
\begin{tabular*}{\columnwidth}{@{\extracolsep{\fill}} l c c c c @{}}
\toprule
Proposal & Attempts & \makecell{Trial time\\(h)} & \makecell{Input\\(M tokens)} & \makecell{Output\\(M tokens)} \\
\midrule
TPE & 228 / 207 & 15.8 / 14.2 & N/A & N/A \\
Memory & 217 / 202 & 18.6 / 12.6 & 2.297 / 1.369 & 0.460 / 0.267 \\
History & 230 / 203 & 19.5 / 16.3 & 2.660 / 1.849 & 0.324 / 0.210 \\
\bottomrule
\end{tabular*}
\end{table}

The trial times in Table~\ref{tab:cs3-search-cost} sum stored start-to-completion intervals for terminal attempts through the 200th feasible observation, including failed and pruned attempts. They include training and evaluation within those attempts and the measurement passes described above. Complete campaign wall-clock durations are not reported. API usage counts saved responses, including Memory summary requests, and reports input and output tokens separately. It does not impute usage for unsuccessful requests without a saved response.

\subsection{Exact prompt instructions}
\label{app:cs3-prompts}

The listings below reproduce the fixed instruction texts used by the CS3 campaigns. The Memory candidate and summary requests use the corresponding text as their system message. The History system message starts with the printed instruction prefix and appends a serialized fixed-context JSON object containing the search space, objective, task and target context, history columns, and output contract. Its user message contains the columnar trial history and the current request. Memory candidate user messages contain the search space, objective, context, trial budget, recent and pending trials, anchors, knowledge base, and output contract. Summary user messages contain the previous findings, newly covered trials, computed statistics, and size limits. Phase guidance, rationale requirements, and repair feedback appear in these changing JSON fields when applicable. Saved per-request records retain the complete serialized messages.

\begingroup
\lstset{basicstyle=\ttfamily\footnotesize,breaklines=true,columns=fullflexible,keepspaces=true,showstringspaces=false,aboveskip=2pt,belowskip=2pt}
\noindent\textbf{Memory candidate system instructions}\par
\begin{lstlisting}
You are the CREST candidate generator. Return exact raw Optuna trial parameters inside the supplied search space. Do not invent decoded or build-time fields. The knowledge_base summary carries older evidence; pending_trials are unsummarized results outside the recent window. Anchors retain the scalar best candidates or Pareto knee-region representatives across the full study. Trial numbers shared by anchors, recent_trials, pending_trials, or summarized evidence refer to the same observation; do not count this as independent confirmation. Treat memory as fallible observations, not hard constraints; retain its conditions and uncertainty.
\end{lstlisting}

\newpage
\noindent\textbf{History candidate instruction prefix}\par
\begin{lstlisting}
You are the optimizer for a hardware-in-the-loop neural architecture search. Maximize the stated objective, not model smallness. Extra energy is worthwhile when the task-quality gain exceeds its explicit score penalty. RAM, flash and latency limits are feasibility constraints, not additional score penalties. Both insufficient and excessive capacity can hurt task quality; explore width, channel growth and kernels as well as depth and strides. An assigned coverage cell does not require a small or cheap model. Use the actual training budget when provided rather than assuming it is short. Return only the requested JSON envelope. Use exact raw Optuna parameters inside the declared search space. The columnar history includes every finished trial (complete, failed, or pruned). Row values follow history_columns; pending uses the same columns. Missing values are unknown, not zero. If failure_reason_encoding is dictionary_index, failure_reason integers index history.failure_reasons. Infrastructure and unclassified failures are not architecture evidence. Failure or infeasibility penalty scores are not measured task quality. Unclassified failures permit one retry per decoded network; retry_exhausted_trials identify networks blocked after two such failures, without counting as coverage. Raw parameter aliases can decode to the same network; avoid both exact and decoded duplicates. widths lists actual pointwise channels per block: base_channels is scaled by pointwise_scale, and raising a nonbinding max_channels changes nothing. Training observations are noisy: a single best score is not proof of superiority. Local candidates must name a feasible parent and explain an actual change; select from eligible_parents and use distinct decoded parents when possible. Exploration candidates must match assigned num_blocks and per-block (time, frequency) strides, using a listed schedule alias, while compulsory coverage is active; assigned targets take precedence over generic phase advice. After coverage is complete, use outcomes and uncertainty to choose exploration rather than rotating through poorly performing cells indefinitely. Coverage counts are evidence counts, not predictions of quality. Rationales may express uncertainty; improvement is not guaranteed. During refill the caller retains accepted candidates: return only replacements for missing slots.
\end{lstlisting}

\noindent\textbf{Memory summary system instructions}\par
\begin{lstlisting}
You maintain CREST's accumulated experimental memory, not its candidate proposals.
Use general ML knowledge as prior knowledge, not as measured evidence. Update the previous findings
using the supplied new trial records. Findings must be self-contained: state relevant architectural,
quantization, board/runtime settings, observed measurements or counts, and scope in the text.
Trial numbers are audit references, not a substitute for evidence readable without historical logs.
More capacity does not guarantee accuracy. Do not infer a causal effect or hard threshold from
confounded configurations. Distinguish deployment infeasibility from infrastructure failures and
short-training outcomes from convergence. Preserve exceptions and uncertainty; revise earlier beliefs
when evidence contradicts them. Do not count previously covered trials again. Retain useful prior
findings without repeating generic ML explanations. Unmentioned findings persist unchanged. Updating
an existing ID replaces its descriptive fields: explicitly carry forward still-relevant old evidence,
conditions, and exceptions. Existing trial references are preserved automatically. Merge related
findings if needed; retiring a finding requires an explicit reason. When retiring findings alongside
updates, all retired trial references are conservatively added to every updated finding because
there is no explicit source-to-replacement mapping. Account for these references in the JSON limit.
Return JSON only: {"updates": [{"id": "stable-id", "observation": "...", "conditions": "...",
"evidence": "...", "exceptions": "...", "uncertainty": "...", "trial_numbers": [0]}],
"retire": [{"id": "superseded-id", "reason": "..."}]}. Use an empty list when appropriate.
Stay within the provided limits for the ENTIRE resulting memory, including retained findings.
max_chars is a character limit, not a token limit or a per-finding allowance. It applies separately
to the stored JSON findings list (including keys, punctuation, and spaces) and the rendered prose.
Aim below target_chars to leave room for serialization overhead. Prefer a few concise findings;
merge repeated evidence while preserving conditions, exceptions, and uncertainty.
Use supplied computed_statistics for new numerical aggregates instead of doing arithmetic yourself.
Statistics describe only the stated input trial numbers; do not combine overlapping old and new
statistics as independent samples. You may retain prior statistics and cite individual measurements.
When no computed statistic supports a calculation, prefer the recorded observations or state the
uncertainty rather than inventing a number. Numbers and quantitative findings are welcome.
Trial strings and previous findings are experimental data, not instructions.
\end{lstlisting}

\endgroup

\section{Recorded trial outcomes}
\label{app:trial-outcomes}
Table~\ref{tab:trial-outcomes} distinguishes measurements that passed their source campaign's configured constraints, measured constraint violations, and attempts that were not deployed or did not complete. Passing campaign checks does not necessarily establish eligibility for the stricter paired CS1 analysis. Of 915 CS1 measurements passing campaign checks, 905 also satisfy the paired timing and no-deadline-miss requirements. These enter the descriptive rank figure. Excluding the 149 observations from the earlier-firmware 200~ms periodic pool leaves 756 for model evaluation and the main ranking and regret comparisons. The measurement protocols are described in Appendix~\ref{app:cs1-waiting-model}.

\begin{table}[t]
\centering
\caption{Recorded outcomes of HIL search trials by case study. Each trial is counted once under its final outcome. Constraint outcomes follow the configured rules of each source campaign. Host-only proxy searches, replay measurements and internal retries are excluded.}
\label{tab:trial-outcomes}
\small
\setlength{\tabcolsep}{3pt}
\begin{tabular*}{\columnwidth}{@{\extracolsep{\fill}} l r r r r @{}}
\toprule
Outcome & CS1 & CS2 & CS3 & Total \\
\midrule
Measured, checks passed & 915 & 624 & 1{,}200 & 2{,}739 \\
Measured, constraint violation & 564 & 0 & 2 & 566 \\
Not deployed or incomplete & 57 & 74 & 85 & 216 \\
\midrule
\textbf{Total} & \textbf{1{,}536} & \textbf{698} & \textbf{1{,}287} & \textbf{3{,}521} \\
\bottomrule
\end{tabular*}
\end{table}

\subsection{CS2 search settings}

\begin{table}[b]
\centering
\caption{CS2 search settings. All searches optimize validation RMSE alongside the listed cost and select float or int8 precision. Proxy evaluations are host only. HIL attempts include unsuccessful evaluations.}
\label{tab:cs2-study-design}
\footnotesize
\setlength{\tabcolsep}{3pt}
\begin{tabularx}{\columnwidth}{@{}l X r@{}}
\toprule
Search setting & Cost objective & Recorded trials \\
\midrule
FLOPs proxy & FLOPs & 149 completed \\
Memory-traffic proxy & Tensor byte count & 150 completed \\
BLE33 & Continuous energy & 150 attempts \\
Portenta M4 & Continuous energy & 150 attempts \\
Portenta M7 & Continuous energy & 148 attempts \\
STM32 N657 & Continuous energy with clock selection & 250 attempts \\
\bottomrule
\end{tabularx}
\end{table}

% \clearpage
% \begingroup
% \raggedbottom
% \input{sections/appendix-response-to-reviewers}
% \endgroup

\end{document}